\documentclass[conference,compsoc]{IEEEtran}

\pdftrailerid{}

\ifCLASSOPTIONcompsoc
  \usepackage[nocompress]{cite}
\else
  \usepackage{cite}
\fi
\usepackage{amsmath,amssymb,amsthm}
\usepackage{graphicx}
\usepackage{booktabs}
\usepackage{array}
\usepackage{xcolor}
\usepackage[hyphens]{url}
\usepackage[hidelinks]{hyperref}
\usepackage{tikz}
\definecolor{sphit}{HTML}{D55E00}
\definecolor{spmiss}{HTML}{0072B2}
\definecolor{spcat1}{HTML}{009E73}
\definecolor{spcat2}{HTML}{882255}
\definecolor{spink}{HTML}{000000}
\definecolor{spgray}{HTML}{999999}
\definecolor{spgraylight}{HTML}{D9D9D9}
\definecolor{spgraypale}{HTML}{F2F2F2}
\definecolor{spwhite}{HTML}{FFFFFF}
\definecolor{sphittint}{HTML}{F4D7BF}
\definecolor{spmisstint}{HTML}{BFDCEC}
\definecolor{spcat1tint}{HTML}{BFE7DC}
\definecolor{spcat2tint}{HTML}{E1C8D4}
\definecolor{spleak1}{HTML}{F7DFCC}
\definecolor{spleak2}{HTML}{ECB78C}
\definecolor{spleak3}{HTML}{E18B47}
\definecolor{spleak4}{HTML}{D55E00}
\definecolor{spleak5}{HTML}{8A3D00}

\usetikzlibrary{arrows.meta,calc,positioning,patterns,decorations.pathreplacing}
\tikzset{
  sp/fig/.style={font=\footnotesize, line width=0.6pt, line cap=butt,
    line join=miter, inner sep=1.5pt, outer sep=0pt,
    >={Stealth[length=3pt,width=2.4pt]}},
  sp/thin/.style={line width=0.4pt},
  sp/thick/.style={line width=1pt},
  sp/step/.style={draw=spink, sp/thin, fill=spwhite, rectangle},
  sp/req/.style={draw=spink, sp/thin, fill=spwhite, rectangle},
  sp/other/.style={draw=spink, sp/thin, fill=spgraylight, rectangle},
  sp/hit/.style={draw=sphit, sp/thin, fill=sphittint, rectangle},
  sp/miss/.style={draw=spmiss, sp/thin, fill=spmisstint, rectangle},
  sp/saved/.style={draw=sphit, sp/thin, dashed, dash pattern=on 1.5pt off 1pt,
    pattern=north east lines, pattern color=sphit, rectangle},
  sp/shift/.style={draw=sphit, sp/thick, {Bar[width=4pt]}-{Stealth[length=3pt,width=2.4pt]}},
  sp/arrow/.style={draw=spink, sp/thin, ->},
  sp/time/.style={draw=spink, sp/thin, ->},
  sp/brace/.style={draw=spink, sp/thin, decorate,
    decoration={brace, amplitude=2.5pt, raise=1pt}},
  sp/ref/.style={draw=spgray, sp/thin, dashed, dash pattern=on 2pt off 1.5pt},
  sp/label/.style={inner sep=1pt, text=spink},
  sp/world/.style={inner sep=1pt, anchor=east, text=spink},
}

\hypersetup{pdftitle={Where the Savings Go: The Limits of Hiding Cross-Tenant Cache Hits in LLM Serving},
  pdfauthor={}, pdfsubject={}, pdfkeywords={}, pdfcreator={}}

\newtheorem{theorem}{Theorem}
\newtheorem{lemma}{Lemma}
\newtheorem{proposition}{Proposition}

\theoremstyle{definition}
\newtheorem{definition}{Definition}

\newcommand{\tablenote}[1]{\par\vspace{3pt}\parbox{\linewidth}{#1}}

\begin{document}

\title{Where the Savings Go: The Limits of Hiding\\
Cross-Tenant Cache Hits in LLM Serving}

\author{\IEEEauthorblockN{Bowen Sun, Chaowei Xiao}
\IEEEauthorblockA{Johns Hopkins University, Baltimore, USA\\{bsun39,chaoweixiao}@jhu.edu}}

\maketitle

\begin{abstract}
LLM serving engines cache prompt prefixes; a cache hit skips their computation
and returns faster, so a cache shared across tenants reveals whether others have
sent a given text. Per-account salting closes this channel but forgoes cross-
tenant reuse, and isolates only whole accounts, even where access is
redistributed or trust domains are subdivided below them. Defenses that share
globally but hide hits promise reuse without relying on account boundaries. How
far can they go? Hiding a hit requires everything tenants observe afterwards to
be distributed as if it had missed. The defenses we examine target the gap
between a hit request and a miss request; we instead follow where the saved time
goes. Step-based schedulers that never idle while work waits pass it in full to
every request queued behind the hit, shifting each earlier by the same amount.
An attacker within ordinary rate limits brackets a probe with two groups of its
own listener requests that each return one token, sends them in one short burst
that other traffic rarely interrupts, and counts in how many steps the listeners finish. Replaying a production trace
on one vLLM instance at up to half its capacity, a single probe tells whether
a text of 256 tokens or more was cached with advantage of at least 0.98. Hiding only the hit request falls
to one probe on vLLM and SGLang; random delays must reach about ten times the shift or more to cut the advantage to 0.1, far above the noise evaluated in prior work;
narrowing what is shared gives no guarantee against this inference. Since any
benefit a tenant can observe is also a signal, we prove that a rate-limited
tenant's gain from content written by $k$ other accounts is at most of order $k$
times the leakage allowed, whatever the provider does with identities; at zero leakage, only the provider
keeps the savings. Exact hiding is attainable in principle on one instance, but
a mechanism that only pauses or holds work knows how long skipped work would
have taken only approximately, so under saturated prefill it must pad every step
that may hold a hit, at a price that grows with time rather than with the number
of hits.
\end{abstract}

\IEEEpeerreviewmaketitle

\section{Introduction}\label{sec:intro}

LLM serving engines keep the key-value tensors computed for a prompt, the KV cache, and reuse them when a later request starts with the same prefix; such a cache hit skips the prefill computation of that prefix, so the response starts sooner~\cite{vllm2026automatic,openai2026prompt,anthropic2026prompt}.
If tenants share one cache, this speed reveals whether someone else has recently sent a given text, and a line of attacks detects such sharing or recovers prompts this way~\cite{wu2025know,song2025early,zheng2026inputsnatch,gu2025auditing,wang2026optileak,sun2026agent}.
Commercial APIs therefore scope their caches to an organization, workspace or account~\cite{openai2026prompt,anthropic2026prompt,fireworks2026prompt}, and vLLM offers a per-request salt on cache keys; Fireworks and vLLM name timing attacks as the reason~\cite{fireworks2026prompt,vllm2026security}.
Such salting forgoes reuse across tenants, and its isolation ends at the account.
Yet access is often redistributed below the account: relays, resellers and applications serve many end users under one account~\cite{zhang2026real,lin2026behavioral}, and we and others have observed relays that merge their clients' identities in production~\cite{sun2026keypooling,fahey2026cacheprobe}.
Within one account, too, teams, end users or sessions may need isolation from each other.
Isolating them requires the distributor to pass each boundary down, and every finer level gives up more reuse.
Defenses proposed in research instead keep a shared cache and hide its hits, promising reuse without relying on account boundaries.
How far can they go?

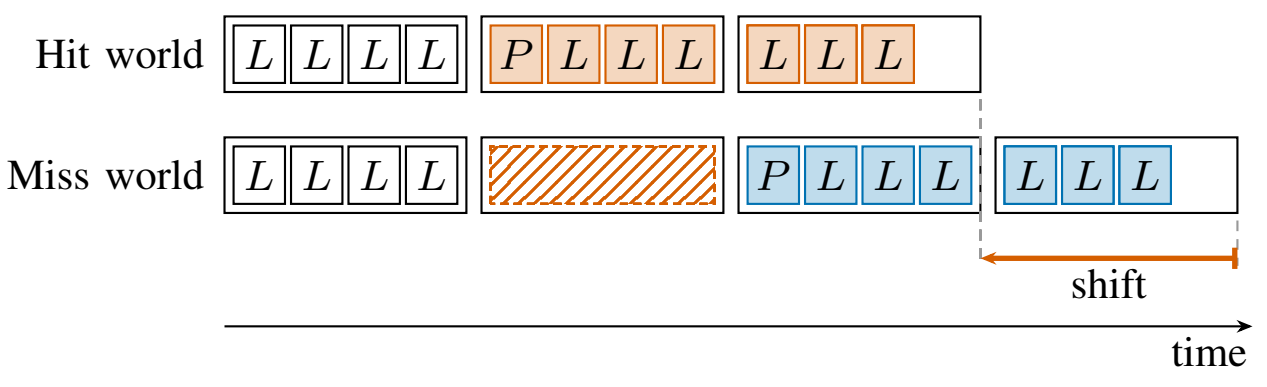
\begin{figure}[t]
\centering
\begin{tikzpicture}[sp/fig, x=1mm, y=1mm]
  \newcommand{\slot}[5]{\draw[#1] (#2+0.6+#3*3.8,#4+0.6) rectangle ++(3.4,3.8) node[midway] {#5};}
  \node[sp/world] at (-1,2.5) {Hit world};
  \node[sp/world] at (-1,-5.5) {Miss world};
  \foreach \x in {0,17,34} \draw[sp/step] (\x,0) rectangle ++(16,5);
  \foreach \x in {0,17,34,51} \draw[sp/step] (\x,-8) rectangle ++(16,5);
  \foreach \i in {0,...,3} {\slot{sp/req}{0}{\i}{0}{$L$} \slot{sp/req}{0}{\i}{-8}{$L$}}
  \slot{sp/hit}{17}{0}{0}{$P$}
  \foreach \i in {1,2,3} \slot{sp/hit}{17}{\i}{0}{$L$}
  \foreach \i in {0,1,2} \slot{sp/hit}{34}{\i}{0}{$L$}
  \draw[sp/saved] (17.6,-7.4) rectangle ++(14.8,3.8);
  \slot{sp/miss}{34}{0}{-8}{$P$}
  \foreach \i in {1,2,3} \slot{sp/miss}{34}{\i}{-8}{$L$}
  \foreach \i in {0,1,2} \slot{sp/miss}{51}{\i}{-8}{$L$}
  \draw[sp/ref] (50,-0.5) -- (50,-11.5);
  \draw[sp/ref] (67,-8.5) -- (67,-11.5);
  \draw[sp/shift] (67,-11) -- (50,-11) node[midway, below=1pt, sp/label] {shift};
  \draw[sp/time] (0,-15.5) -- (68,-15.5) node[pos=1, below left=1pt and 0pt, sp/label] {time};
\end{tikzpicture}
\caption{Where a hit's saving goes.
Frames are engine steps, each running queued requests in arrival order up to a budget of work, here four illustrative units; $L$ marks the attacker's listener requests, each with one output token, before and after its probe~$P$ for a candidate text.
In the miss world $P$ also computes its prefix, hatched; in the hit world another tenant has recently sent that prefix, so it is cached and skipped.
$P$ and every request behind it, tinted, then finish one step earlier; the dashed lines mark when the last listener finishes, and the gap between them, the shift, is what the listeners read by counting steps.}
\label{fig:overview}
\end{figure}

To hide a hit, it is not enough that the hit request itself looks like a miss.
As noninterference requires~\cite{goguen1982security}, everything tenants observe from then on must be distributed in the world where an attacker's request for a given text, the probe, hits as in the world where it misses: when each token of every later request arrives, its content, usage and billing, and any rejection.
The defenses we found all evaluate the leak as the gap between a hit request and a miss request; some mention interference among concurrent requests or define security over all latency the attacker observes, but none analyzes or evaluates how the resources a hit saves affect later requests, and Appendix~\ref{app:defense-tm} quotes their threat models.
We instead follow where the saved time goes.

It goes to the requests behind the hit.
Mainstream engines schedule in steps~\cite{yu2022orca,agrawal2024taming}: each step runs a batch of requests within a per-step token budget, Table~\ref{tab:engines}, and, in their code, the scheduler is work-conserving: it never idles while work waits.
A hit removes part of the prefill from this sequence, so every request queued behind it starts and finishes earlier by the same amount, which we call the shift; the saving is copied to each of them, not divided among them.
On an idle vLLM engine on an L40S GPU, a 1,024-token hit makes all 16 requests queued behind it finish 68 to 69\,ms earlier, and an attacker that sizes its requests to the step budget turns even a 16-token hit into a shift of a whole step for all of them but the last.
The attack in Figure~\ref{fig:overview} brackets a probe for a candidate text with two groups of the attacker's own listener requests, each with one output token, sent with the probe in one short burst, and counts in how many steps the listeners finish.
Counting steps makes the reading insensitive to queueing delay; its main condition is that no other tenant's request enters the short window around the probe, which Section~\ref{sec:threat} turns into a required burst rate.
Prior attacks read the probe's own response: its time to first token~\cite{song2025early,zheng2026inputsnatch,gu2025auditing}, a step-wave pattern in that time~\cite{sun2026agent}, or, in the construction closest to ours, its place among dummy requests sent before and after it under longest-prefix-match scheduling~\cite{wu2025know,wang2026optileak}.
Reading another party's work through the timing of one's own queued requests is classical~\cite{vila2017loophole,tramer2020remote,beams2021packet}, and work-conserving schedulers are known to leak load across users~\cite{kadloor2013delay,gong2016quantifying}.
New here is what is read: a saving that another tenant's cached text gives the probe, which the engine copies to every queued request, so listeners that never touch the text read it, and hiding the probe's own timing does not help.

\begin{table}[t]
\centering
\caption{Schedulers of mainstream serving engines}
\label{tab:engines}
\footnotesize
\setlength{\tabcolsep}{4pt}
\begin{tabular}{@{}lll@{}}
\toprule
Engine & Budget & Prefill and decode \\
\midrule
vLLM & 2,048 to 16,384 by memory & Mixed \\
SGLang & 2,048 and up by memory & Separate \\
TensorRT-LLM & 8,192 & Mixed \\
TGI & Automatic & Mixed if chunked \\
LMDeploy TurboMind & 8,192 & Mixed \\
LMDeploy PyTorch & 8,192 & Separate \\
DeepSpeed-FastGen & 768 & Mixed \\
llama.cpp server & 2,048 & Mixed \\
MLC-LLM & Model configuration & Mixed \\
LightLLM & 16,384 split over ranks & Separate \\
\bottomrule
\end{tabular}
\tablenote{Read from each engine's code and documentation.
Every engine schedules in steps, each selecting a batch, running one forward pass and releasing its outputs, and by default admits waiting requests in arrival order.
Budget: default tokens per step; mixed: prefill and decode share a step.}
\end{table}
The attacker is weak: a few ordinary accounts within their rate limits, with no control or knowledge of other tenants' traffic and no assumption about how the provider uses identities.
It asks whether a given text is in the cache, the standard membership question~\cite{carlini2022membership}: if the provider reads content only through cache lookups, security against membership inference bounds every inference over block-aligned candidates of equal length, as Section~\ref{sec:threat} shows.
We report the one-shot advantage, the probability of answering ``cached'' when the text is cached minus that when it is not~\cite{yeom2018privacy}, with one probe per secret.
Four requests suffice on an idle vLLM engine, with advantage 0.985 to 0.989; replaying a public production trace~\cite{patel2024splitwise} on one instance at up to half of its capacity, a single probe still reaches advantage of at least 0.98, also when repeated with every parameter frozen.
Secrets of 4,096 tokens or more stay readable, with advantage 0.66 to 1.00, behind a production router over two and four cache-affinity instances, on two other GPU types, and with a 70B model up to a quarter of capacity.
Section~\ref{sec:attack} also reports the limits: the shortest readable secret grows with the per-step token budget, and under routing that ignores the cache a single probe reads only the instance it lands on.

Section~\ref{sec:defenses} organizes published defenses by three knobs: which content may be shared, when each response is released, and how the provider reacts afterwards.
Defenses that make only the hit request look like a miss~\cite{gu2025auditing,zheng2026inputsnatch,wu2025know} leave the copied saving in place, and the listeners read it with one probe on vLLM and on SGLang, with advantage 0.96 to 0.99.
Random delays blur the shift without removing it.
Under the 20\,ms Gaussian noise that KVGov reports to stop a timing model of PromptPeek~\cite{wu2025know} in its simulation~\cite{addagada2026governing}, the listeners keep advantage above 0.98 under load; cutting it to 0.1 on an idle engine takes mean exponential delays of 0.15 to 4.9\,s per response, 9 to 16 times the shift.
Defenses that narrow what is shared or react afterwards~\cite{chu2025selective,wu2026cacheprune,addagada2026governing,song2025early,sun2026agent,pennas2026cachesolidarity,zuo2026privkvsched} give no guarantee for membership: content that stays shareable remains readable, and the per-block owner flags of CacheSolidarity~\cite{pennas2026cachesolidarity} become a new channel, read with advantage 0.97 to 0.98.
Classical defenses that keep the saving inside the system~\cite{askarov2010predictive,zhang2011predictive,aviram2010determinating,li2014stopwatch,oliverio2017secure} either give tenants nothing or, when margins or release times are imprecise, leak through timing, for instance through faster steps after the GPU idles.

These results share one cause: a tenant sees only the observations that noninterference covers, so any benefit it draws from shared content is also a signal.
Extending privacy costs known for caches and deduplication~\cite{armknecht2017side,acs2013cache,ghazi2026on}, we prove that if no rate-limited tenant can tell the two worlds apart, none gains from content that other accounts wrote, and that with leakage~$\varepsilon$ allowed, the gain from content written by $k$ other accounts is at most of order $k\varepsilon$, whatever the provider does with identities.
At zero leakage only the provider keeps the savings, and zero leakage is attainable under premises including a single instance, batch-invariant kernels~\cite{he2025defeating} and exact release timers: the engine runs in lock step with a replica of its scheduler that has no cross-tenant hits~\cite{devriese2010noninterference,li2014stopwatch}, skipping the cached work but releasing each response at the replica's time.
Since the provider knows skipped work's duration only approximately, the replica's steps must last a bound fixed in advance, and every step that may contain a hit is padded to cover that uncertainty, at a cost that grows with time rather than with the number of hits; Section~\ref{sec:defenses} states the premises.
Providers that need isolation should therefore salt per trust domain, with the boundary passed down by the layer that knows it, as Section~\ref{sec:discussion} discusses.

We make four contributions.

\textbf{Attack.} What hiding a hit requires, how step scheduling copies a hit's saving to later requests, and a step-counting attack with its correctness condition, measured across engines, loads, instances, GPUs and models, in Sections~\ref{sec:threat} and~\ref{sec:attack}.

\textbf{Defenses.} A class-by-class evaluation of published defenses along the three knobs, with an attack or a price lower bound for each class, in Section~\ref{sec:defenses}.

\textbf{Limits.} Proofs that a visible benefit is a signal and that the gain from shared content is bounded by the leakage allowed, and a construction attaining zero leakage with its cost, also in Section~\ref{sec:defenses}.

\textbf{Method.} Two-world audits with one-shot advantage that replay defenses on records of an unmodified engine.

\section{System and Threat Model}\label{sec:threat}

\subsection{System Model}\label{sec:system}

An engine stores the KV cache in blocks of a fixed number of tokens, 16 in vLLM.
The key of a block hashes its tokens together with the key of the block before it, and a per-request salt, if any, enters the first key; an arriving request therefore hits the longest cached prefix of its prompt, block by block, and computes the rest.
Blocks stay cached after their request ends until memory is needed, and providers describe retention of about five minutes to 24 hours~\cite{openai2026prompt,anthropic2026prompt}.
A provider tells tenants apart only by account, so we call a hit on a block that another account wrote a cross-tenant hit, and a hit on the account's own block a self-hit.
Deployments may run several instances behind a router, with separate caches or a shared pool.

Mainstream engines schedule in steps~\cite{yu2022orca,agrawal2024taming}.
Each step first serves running requests, a decode taking one token and an unfinished prefill its remaining prompt, and then admits waiting requests first come, first served.
Every request takes the tokens it needs, up to what remains of the per-step token budget~$B$, so at most one prefill is cut, and its remainder runs first in the next step; a step also holds at most $K$ sequences.
The scheduler is work-conserving: the next step starts as soon as the previous one ends while work waits, and requests whose prefill completes in the same step receive their first tokens together; vLLM releases at most about 128 at a time, more than any of our rulers puts in a step, Appendix~\ref{app:experiments}.
With its defaults on an L40S GPU, vLLM~0.26 uses $B = 2{,}048$ and $K = 256$; SGLang~0.5.16 admits first come, first served into prefill batches of a fixed chunk of 4,096 tokens without decodes.
Table~\ref{tab:engines} summarizes the schedulers of mainstream engines; NVIDIA Dynamo hands scheduling to such engines, Ollama runs llama.cpp, and Appendix~\ref{app:engines} says when and how we read them.

We assume that a step's duration depends only on its shape, that is, the number of tokens and sequences and the context length of each, plus noise that does not depend on cached content.
This holds for the dense models we study; mixture-of-experts models route tokens by content, so our theorems do not cover them, though our attack, which counts steps, reads secrets on one, Section~\ref{sec:load}.
It holds only approximately when clock boost after idling makes step durations depend on recent load, which Section~\ref{sec:defenses} exploits.

\subsection{Victim, Attacker and Security}\label{sec:security}

The victim is one account that uses a secret text~$S$, possibly over several requests such as the turns of a conversation.
In the hit world~$H$ it sends~$S$; in the miss world~$M$ it sends instead an equally long text~$S'$ shared with no one, replacing every occurrence, with all else equal.
A probe~$P$ for~$S$ thus hits in~$H$ and misses in~$M$.
The victim meets three conditions: the attacker does not use its account; its output lengths and stop conditions are fixed, as by \texttt{max\_tokens} with \texttt{ignore\_eos}; and the length, parameters and timing of its later requests may depend on when responses arrive and how long they are, but not on their content.
Without these conditions its own behavior reveals the secret even under per-account salts, a content channel outside our scope; with them, per-account salting leaks nothing.

The attacker holds a few ordinary accounts within a rate budget~$\beta$, the provider's rate limits summed over them: in any interval of length~$t$ it sends at most $b_r + rt$ requests with at most $b_t + \rho t$ input tokens, with at most $c$ in flight.
It knows the candidate text and the engine's configuration, which fresh requests can also measure: on an idle engine, the way the first tokens of a burst of its own requests cluster gives the per-step budget, Appendix~\ref{app:experiments}.
It neither controls nor knows other tenants' traffic and observes only its own requests: the time, content and usage fields of each output event, status codes and headers, and its billing records.
The provider's mechanism is any causal randomized algorithm for admission, scheduling, caching, routing and release, and may read all content and use identities.
We assume nothing about how it uses them: the definition below covers every account within the rate budget, and a defense that hides savings only from the account whose request hit is defeated by placing the probe and the listeners on accounts the provider cannot link.
A workload is the set of accounts active in one cache epoch, the period from a given cache state until the content sent in it can no longer be cached, with their programs, content and randomness and that starting state; we call the traffic of accounts other than the attacker's and the victim's honest.

\begin{definition}[Security]\label{def:security}
A mechanism is $(\beta,\varepsilon)$-secure if, for every workload~$W$, every account~$a$ in~$W$ that meets the victim conditions, and every set~$O$ of other accounts within rate budget~$\beta$, the total variation distance between $O$'s views in~$W$ and in~$W^a$ is at most~$\varepsilon$, where $W^a$ replaces every text that $a$ shares with other accounts by fresh random tokens and a view is the joint record of all of $O$'s requests over the epoch.
\end{definition}

The one-shot advantage $\varepsilon^*(\beta)$ replaces a single shared text instead of all, the setting of one membership probe per secret; the best decision rule's advantage, defined in Section~\ref{sec:intro}, equals this distance, and $\varepsilon^*(\beta) = 0$ implies $(\beta,0)$-security.
The worst case over workloads, which includes an idle engine, matters because quiet periods exist and the attacker may wait for one within the retention time; results under honest load measure how realistic the attack is.

Membership is the basic question.
If the mechanism reads content only through cache lookups, unlike content-hash routing, content filters and semantic caches, the two worlds coincide until the first lookup that differs, so security for membership bounds every inference over equal-length candidates.

\begin{proposition}[Membership]\label{prop:membership}
Let the secret be one of several candidates that differ only in at most $d$ block-aligned segments at the same positions.
If the one-shot advantage is at most~$\varepsilon$ for every segment, the attacker's views under any two candidates differ in total variation by at most $T = 2d\varepsilon$, or $T = 2\varepsilon$ under $(\beta,\varepsilon)$-security, and for prior~$\mu$ it guesses the secret with probability at most $\max\mu + (1-\max\mu)\,T$.
\end{proposition}

The proof, in Appendix~\ref{app:proofs}, walks from one candidate to the other a segment at a time through fresh tokens, so at most $2d$ neighboring pairs separate them, each changing the view by at most~$\varepsilon$, and the guessing bound follows as in group privacy.
Membership does not imply extraction, which also needs the victim's blocks cached over many probes.

\subsection{What the Attack Needs}\label{sec:needs}

The attack of Section~\ref{sec:attack} sends a group of listeners just before the probe and another just after it.
Its reading window runs from when the first listener before the probe enters the service to when the last request after the probe does, widened by the jitter of request entry, and the attack needs three conditions in this window, the window conditions.
First, no other request, the victim's included, enters the service in the window.
Second, no request jumps ahead through priorities or a queue of blocked requests.
Third, the engine is at no limit: sequence and memory limits leave room for the probe and the listeners, the victim's blocks are still cached, and few decodes ahead of the listeners finish during the steps that are read.
When these hold, and the listeners have equal length and the timing and separation conditions of Theorem~\ref{thm:ruler} are met, the reading is correct on every path.
The advantage under a given honest load is therefore at least $\Pr_H[G_H] + \Pr_M[G_M] - 1$, where $G_H$ and $G_M$ are the events that all three conditions hold in~$H$ and in~$M$; only the first surely has the same probability in both worlds.

A rate-limited attacker makes the first condition likely by sending the window's requests in a burst.

\begin{proposition}[Burst rate]\label{prop:burst}
Let other requests, the victim's included, arrive as a Poisson process of rate~$\lambda$, and let the attacker send the window's $n$ requests at rate~$r_s$, so that the window lasts $w = w_0 + (n-1)/r_s$, where $w_0$ is its length, entry jitter included, when all $n$ are sent at once.
No other request enters the window with probability $e^{-\lambda w}$, which is at least~$p$ if and only if $r_s \ge (n-1)/\bigl(\ln(1/p)/\lambda - w_0\bigr)$; when $\lambda \ge \ln(1/p)/w_0$, no rate suffices.
\end{proposition}

On an L40S GPU, $w_0$ is about 22\,ms, estimated from our replays; one instance then keeps the window clear with probability $p = 0.8$ only below 10.1 other requests per second.
At 2.27 honest requests per second, half of capacity in Section~\ref{sec:load}, reading a 1,024-token secret takes 51 requests in the window sent at 655 per second, that is, within about 76\,ms, while the attacker's average rate stays low.
The advantage under such load is at least $2(p - \eta) - 1$, where $\eta$ bounds the probability that the other conditions fail; with several instances, $\lambda$ is the arrival rate at the entry, and Appendix~\ref{app:experiments} tabulates other lengths and loads and gives sufficient conditions for other arrival processes.
Providers may enforce per-minute limits over shorter intervals~\cite{openai2026rate}; at one second, a burst of $n$ requests needs a limit of about $60n$ requests per minute, about 3,000 for the 51 requests above, or a few accounts.

One reading, the probe with its listeners, costs 20 to 176 requests for secrets of 4,096 down to 256 tokens and about 17,000 to 22,000 input tokens, with one output token each; on an idle vLLM engine four requests suffice.
Accounts can be reused across probes, so they are a fixed cost.
The reading tolerates jitter in the attacker's timestamps up to about a quarter of a full step, about 38\,ms on an L40S GPU, as a worst-case bound.

A provider might try to detect such bursts.
Suppose it serves a class of rate-limited honest tenants whose traffic looks like the attack's, such as classification jobs that send batches of long documents with \texttt{max\_tokens} of one.
Traffic detection then does not stop an attacker that imitates this class: it sends its requests as such a batch and puts the candidate, followed by fresh tokens, into one of them.
With accounts whose attributes match that class, it differs from those tenants only in content, so any rule that reads timing, lengths, parameters, metadata or accounts but not hits must block those tenants at the same rate, or keep the saving from reaching them, and then they gain nothing from shared content by Section~\ref{sec:defenses}.
This needs knowledge of the class's distributions; classifiers that recognize probes by content are reactive defenses, treated in Section~\ref{sec:defenses}.

\section{Copied Savings: Reading Hits Through Other Requests}\label{sec:attack}

\subsection{Where the Saving Goes}\label{sec:where}

A hit skips the prefill of the cached prefix, and the GPU time and memory it saves, the saving, surface in at least four places, which we call its exits, listed in Table~\ref{tab:exits}.
A defense must keep each of them out of every tenant's view; making the hit request look like a miss closes only the first.
We do not claim the list is complete: the requirement of Section~\ref{sec:threat} is stated on what tenants observe and needs no list.

\begin{table}[t]
\centering
\caption{Where a hit's saving surfaces}
\label{tab:exits}
\footnotesize
\begin{tabular}{@{}p{0.21\columnwidth}p{0.37\columnwidth}p{0.32\columnwidth}@{}}
\toprule
Exit & What changes & What a tenant observes \\
\midrule
The hit request & It finishes sooner, and its bill may show a discount & Its own timing and bill \\
Requests behind it & They start and finish earlier & Timing of its later requests \\
Shared state & Other cached data stays longer, and a defense's own bookkeeping changes & Whether its own blocks were evicted; flags a defense exposes \\
Hardware & The GPU idles, and later steps run faster & Timing of later requests \\
\bottomrule
\end{tabular}
\end{table}

Under step scheduling the second place takes the saving whole.
Lemma~\ref{lem:copy} states this for an engine that holds only the attacker's requests, which is one of the workloads that Definition~\ref{def:security} quantifies over; Appendix~\ref{app:proofs} gives the proof.

\begin{lemma}[Copied saving]\label{lem:copy}
On an engine that holds only the attacker's requests, let a probe hit $h$ tokens, and let it and the requests behind it each output one token, with the sequence limit not binding.
Until the engine first idles after the probe:
(i) by the end of every step the engine has done $h$ more tokens of the attacker's work in~$H$ than in~$M$, so this lead is never used up and no request behind the probe finishes earlier in~$M$;
(ii) if the probe runs alone in its step, every request behind it finishes later in~$M$ by the same amount, the difference between the probe's step durations in the two worlds;
(iii) if the probe and the first request behind it together fill one step's budget~$B$ and every later request has $B$ tokens, every one of them finishes exactly one step later in~$M$, for any $1 \le h \le B$.
\end{lemma}

\begin{figure}[t]
\centering
\includegraphics{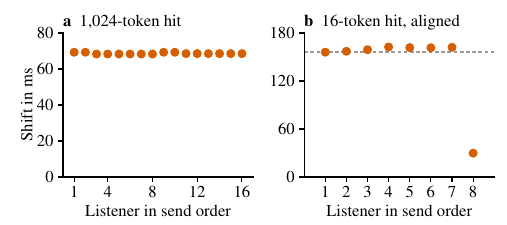}
\caption{The saving is copied to every request behind the hit.
Idle vLLM on an L40S GPU; each point is one listener's shift, its mean completion time in the miss world minus that in the hit world, over 140 to 154 trials per world, with 95\% intervals smaller than the markers.
a: 16 listeners of 256 tokens behind a probe whose 1,024-token prefix hits.
b: eight listeners behind a hit of one 16-token block, sized so that the probe and the first listener fill one step and each later listener a whole step; dashed: one full step, 156\,ms, by which every listener but the last shifts.}
\label{fig:copy}
\end{figure}

All three items follow from first-come-first-served packing: skipping $h$ tokens moves every later token of the attacker's queue $h$ positions earlier.
Item~(i) is why the saving is copied rather than divided; item~(iii) turns any hit up to~$B$ into one more step for every request behind it, a full one for all but the last.
Figure~\ref{fig:copy} shows items~(i) and~(iii) on an idle engine; unless stated otherwise, experiments use vLLM~0.26 with its default scheduler, Llama-3.1-8B-Instruct and one L40S GPU.
With a 1,024-token hit, each of the 16 listeners sent behind the probe finishes 68 to 69\,ms earlier in~$H$.
With a 16-token hit and listeners aligned to the budget, which we call the aligned construction, the first seven of eight listeners finish 156 to 162\,ms earlier, a full step, and the last one about 30\,ms earlier.
With this hit, the probe's own time to first token barely moves: a decision rule on it reaches advantage 0.07, against 0.987 on the listeners, and 0.16 against 0.980 with Qwen3-8B.
SGLang copies the saving likewise, shifting its listeners by 9 to 389\,ms for hits of 256 to 4,096 tokens, read with advantage 0.957 to 0.983.
Decodes already running carry it too: 16 streams decoding when the probe arrives emit their next six tokens 8 to 279\,ms sooner in~$H$ for hits of 256 to 4,096 tokens, read with advantage 0.985 to 0.986.
Advantages here are one-shot, with one probe per secret and fresh content per trial; Appendix~\ref{app:experiments} gives sample sizes and intervals.

\subsection{The Ruler}\label{sec:ruler}

On an idle engine, item~(iii) is already an attack with four requests, each with one output token: a head of $2B$ fresh tokens, the probe, that is, the candidate followed by $u$ fresh tokens, a listener of $B-u$ tokens and a listener of $B$ tokens, sent 3\,ms apart.
The head fills two steps, the probe and the first listener share the third, and the first listener finishes one step later in~$M$ for any block-aligned hit.
Measured from the head's first token, the first listener's first token arrives after about 153\,ms in~$H$ and 304 to 309\,ms in~$M$; for hits of 16, 256, 1,024 and 2,048 tokens, with 133 to 166 trials per world, the advantage is 0.985 to 0.989.

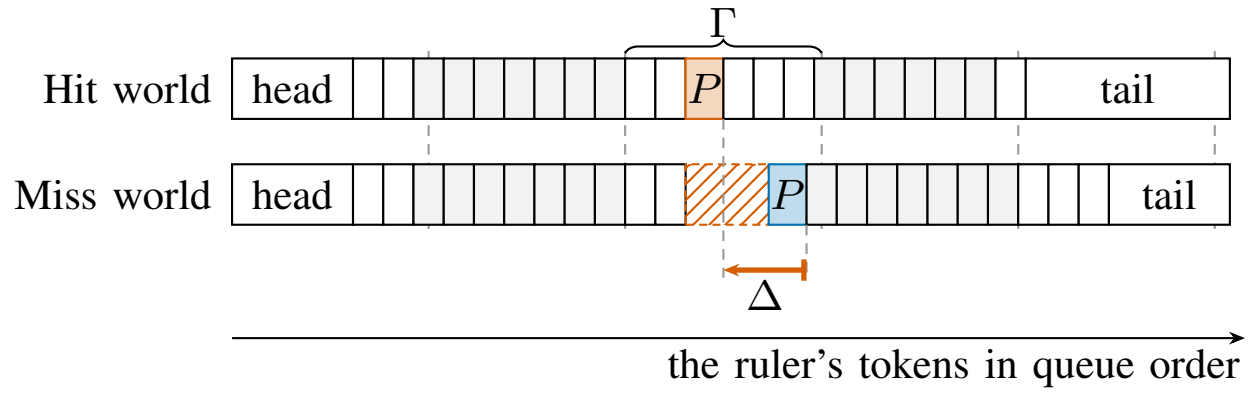
\begin{figure}[t]
\centering
\begin{tikzpicture}[sp/fig, x=1mm, y=1mm,
  sp/rulerpale/.style={draw=spink, sp/thin, fill=spgraypale, rectangle}]
  \node[sp/world] at (-1,2.0) {Hit world};
  \node[sp/world] at (-1,-5.0) {Miss world};
  \draw[sp/ref] (13.0,4.5) -- (13.0,-7.5);
  \draw[sp/ref] (26.0,4.5) -- (26.0,-7.5);
  \draw[sp/ref] (39.0,4.5) -- (39.0,-7.5);
  \draw[sp/ref] (52.0,4.5) -- (52.0,-7.5);
  \draw[sp/ref] (65.0,4.5) -- (65.0,-7.5);
  \draw[sp/req] (0.00,0.00) rectangle (8.00,4.00) node[midway] {head};
  \draw[sp/req] (8.00,0.00) rectangle (10.00,4.00);
  \draw[sp/req] (10.00,0.00) rectangle (12.00,4.00);
  \draw[sp/rulerpale] (12.00,0.00) rectangle (14.00,4.00);
  \draw[sp/rulerpale] (14.00,0.00) rectangle (16.00,4.00);
  \draw[sp/rulerpale] (16.00,0.00) rectangle (18.00,4.00);
  \draw[sp/rulerpale] (18.00,0.00) rectangle (20.00,4.00);
  \draw[sp/rulerpale] (20.00,0.00) rectangle (22.00,4.00);
  \draw[sp/rulerpale] (22.00,0.00) rectangle (24.00,4.00);
  \draw[sp/rulerpale] (24.00,0.00) rectangle (26.00,4.00);
  \draw[sp/req] (26.00,0.00) rectangle (28.00,4.00);
  \draw[sp/req] (28.00,0.00) rectangle (30.00,4.00);
  \draw[sp/hit] (30.00,0.00) rectangle (32.50,4.00) node[midway] {$P$};
  \draw[sp/req] (32.50,0.00) rectangle (34.50,4.00);
  \draw[sp/req] (34.50,0.00) rectangle (36.50,4.00);
  \draw[sp/req] (36.50,0.00) rectangle (38.50,4.00);
  \draw[sp/rulerpale] (38.50,0.00) rectangle (40.50,4.00);
  \draw[sp/rulerpale] (40.50,0.00) rectangle (42.50,4.00);
  \draw[sp/rulerpale] (42.50,0.00) rectangle (44.50,4.00);
  \draw[sp/rulerpale] (44.50,0.00) rectangle (46.50,4.00);
  \draw[sp/rulerpale] (46.50,0.00) rectangle (48.50,4.00);
  \draw[sp/rulerpale] (48.50,0.00) rectangle (50.50,4.00);
  \draw[sp/req] (50.50,0.00) rectangle (52.50,4.00);
  \draw[sp/req] (52.50,0.00) rectangle (66.00,4.00) node[midway] {tail};
  \draw[sp/req] (0.00,-7.00) rectangle (8.00,-3.00) node[midway] {head};
  \draw[sp/req] (8.00,-7.00) rectangle (10.00,-3.00);
  \draw[sp/req] (10.00,-7.00) rectangle (12.00,-3.00);
  \draw[sp/rulerpale] (12.00,-7.00) rectangle (14.00,-3.00);
  \draw[sp/rulerpale] (14.00,-7.00) rectangle (16.00,-3.00);
  \draw[sp/rulerpale] (16.00,-7.00) rectangle (18.00,-3.00);
  \draw[sp/rulerpale] (18.00,-7.00) rectangle (20.00,-3.00);
  \draw[sp/rulerpale] (20.00,-7.00) rectangle (22.00,-3.00);
  \draw[sp/rulerpale] (22.00,-7.00) rectangle (24.00,-3.00);
  \draw[sp/rulerpale] (24.00,-7.00) rectangle (26.00,-3.00);
  \draw[sp/req] (26.00,-7.00) rectangle (28.00,-3.00);
  \draw[sp/req] (28.00,-7.00) rectangle (30.00,-3.00);
  \draw[sp/saved] (30.00,-7.00) rectangle (35.50,-3.00);
  \draw[sp/miss] (35.50,-7.00) rectangle (38.00,-3.00) node[midway] {$P$};
  \draw[sp/rulerpale] (38.00,-7.00) rectangle (40.00,-3.00);
  \draw[sp/rulerpale] (40.00,-7.00) rectangle (42.00,-3.00);
  \draw[sp/rulerpale] (42.00,-7.00) rectangle (44.00,-3.00);
  \draw[sp/rulerpale] (44.00,-7.00) rectangle (46.00,-3.00);
  \draw[sp/rulerpale] (46.00,-7.00) rectangle (48.00,-3.00);
  \draw[sp/rulerpale] (48.00,-7.00) rectangle (50.00,-3.00);
  \draw[sp/rulerpale] (50.00,-7.00) rectangle (52.00,-3.00);
  \draw[sp/req] (52.00,-7.00) rectangle (54.00,-3.00);
  \draw[sp/req] (54.00,-7.00) rectangle (56.00,-3.00);
  \draw[sp/req] (56.00,-7.00) rectangle (58.00,-3.00);
  \draw[sp/req] (58.00,-7.00) rectangle (66.00,-3.00) node[midway] {tail};
  \draw[sp/brace] (26.0,4.0) -- (39.0,4.0) node[midway, above=3pt, sp/label] {$\Gamma$};
  \draw[sp/ref] (32.50,0.0) -- (32.50,-10.5);
  \draw[sp/ref] (38.00,-7.0) -- (38.00,-10.5);
  \draw[sp/shift] (38.00,-10.0) -- (32.50,-10.0) node[midway, below=1pt, sp/label] {$\Delta$};
  \draw[sp/time] (0,-14.5) -- (67.0,-14.5) node[pos=1, below left=1pt and 0pt, sp/label] {the ruler's tokens in queue order};
\end{tikzpicture}
\caption{How the ruler reads a hit.
Its requests in queue order, at an illustrative scale in tokens: a head, listeners before the probe~$P$, the probe, listeners behind it and a tail; dashed lines are step boundaries, each step holding equally many of the ruler's tokens.
Listeners that finish in the same step, shaded alike, get their first tokens together, so the attacker sees where the boundaries fall among its own requests.
In the miss world the probe also computes its prefix, hatched, so every listener behind it sits $\Delta$ further on.
$\Gamma$ counts the attacker's tokens from the last boundary before the probe to the first after it as if the probe hit: one step in~$H$, $\Delta$ less in~$M$.}
\label{fig:ruler}
\end{figure}

Other tenants' requests change what each step holds, so under load absolute times no longer separate the worlds.
Our construction for loaded engines, which we call the ruler, instead reads where step boundaries fall, as Figure~\ref{fig:ruler} illustrates.
It sends, in order, a head of fresh tokens, a group of listeners before the probe, the probe, a group after it and a tail of at least $B$ tokens.
Listeners have $\ell$ fresh tokens and every request outputs one token; the whole ruler goes out as one burst, its groups a few milliseconds apart so that they enter the queue in this order.
Listeners that finish in the same step receive their first tokens together, so the attacker groups their responses by arrival into steps and learns how many of its own tokens each step held.
The listeners before the probe give the ruler's share of each step and the phase of the step boundaries in the attacker's own token count; the first boundary among the listeners after the probe falls $\Delta$ of the attacker's tokens earlier in~$M$, where $\Delta$ is the work the hit skips, $h$ without a defense.
The attacker compares that boundary with the phase predicted for each world, or, when the gaps between groups let it count steps, counts the steps between the last boundary before the probe and the first after it.

\begin{theorem}[Ruler]\label{thm:ruler}
Let $\hat a$ be the attacker's estimate of the tokens it receives per step, from the last $m+1$ boundaries before the probe; $\Gamma$ the distance in its own tokens from the last of them to the first boundary after the probe, counted as if the probe hit; and $F^*$ a tolerance for the decodes of other requests that finish during the read.
Let $\tau = F^* + \ell/m$, $n_{\max} = 1 + \lceil (x_M + \ell)/(\hat a - \tau) \rceil$ with $x_M$ the probe's tokens to compute in~$M$, and $I_n = \bigl(n(\hat a - \tau) - \ell,\ n(\hat a + \tau) + \ell\bigr)$.
Suppose the window conditions of Section~\ref{sec:needs} hold, the head, groups and tail are long enough to keep every step read full, at most $F^*$ of those decodes finish, and each of the attacker's first-token timestamps lies within a quarter of the gap between full steps from the release of its step.
(i) $\Gamma$ lies in some $I_n$ with $n \le n_{\max}$ in~$H$, and in some $I_n - \Delta$ in~$M$.
(ii) If $|\Delta - j\hat a| \ge 2n_{\max}\tau + 2\ell$ for every integer $|j| < n_{\max}$, the unions of the $I_n$ and of the $I_n - \Delta$ are disjoint, and the rule that picks the world whose union holds~$\Gamma$ is correct on every path, without the probe's own response and without absolute delays.
(iii) If the attacker can also count the steps between the two boundaries, $\Delta \ge 2(n_{\max}\tau + \ell)$ suffices.
\end{theorem}

Each change from one group of released responses to the next places a step boundary within $\ell$ of the attacker's tokens, so the steps between the two boundaries hold, up to $\ell$ at each end, $\Gamma$ of its tokens in~$H$ and $\Gamma + \Delta$ in~$M$, and the separation keeps the worlds apart.
When $\Gamma$ falls in neither union, we score the undecided reading as a random guess.
The proof, in Appendix~\ref{app:proofs}, uses only first-come-first-served admission under a per-step budget that does not shrink during the read and release of first tokens per step, so it also covers SGLang's default scheduler.
Other tenants' queueing and request lengths enter only through the window conditions, the decodes that finish during the read and the running sequences that reduce the budget.
The separation needs listeners several times shorter than~$\Delta$, while every step must be filled by about $B/\ell$ of them, so the number of requests grows roughly as $B/h$.
A hit longer than~$B$, such as a 4,096-token secret, is read by counting steps, item~(iii).
The rulers we ran have 20 requests with listeners of 560 tokens for 4,096-token secrets, 52 with 176 tokens for 1,024 and 176 with 48 tokens for 256; all but the head fall in the window.
With $B = 2{,}048$, our listener lengths meet the separation for hits from about 144 tokens; Section~\ref{sec:limits} gives the limit for other lengths.

\subsection{Under Honest Load}\label{sec:load}

We replay the Azure LLM inference trace of 2023~\cite{patel2024splitwise}, conversation part, with its context and output lengths, as other tenants' traffic, at fractions of the rate at which the engine saturates, which equals the trace's original rate.
Half of capacity is 2.27 to 2.93 honest requests per second with a median of 39 to 53 running requests, depending on the trace segment.
Each trial is one cache epoch: the victim sends the secret once, and the ruler follows.
We ran the replay three times. The first run's rulers lacked the tail, so the step after the back boundary could be short and its first tokens merge with the previous step's, most often without load; they read 256-token secrets with only 0.50, 0.87 and 0.94 without load and at a quarter and half of capacity.
We fixed the second run's design before collecting its data, and the third repeats it with every parameter frozen, fresh seeds and a level without load.

\begin{figure}[t]
\centering
\includegraphics{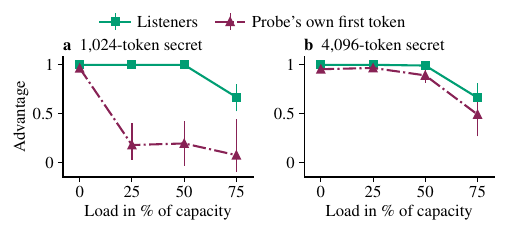}
\caption{Under honest load, listeners read 1,024-token secrets that the probe's own timing no longer shows.
vLLM on an L40S GPU with the Azure conversation trace at fractions of the rate that saturates the engine; the second run of Section~\ref{sec:load}, with the points at no load from the third, its frozen repetition.
One-shot advantage of the ruler's decision rule, and of a threshold on the probe's own time to first token fitted on half of the trials and scored on the rest; bars: 95\% bootstrap intervals over 30 to 66 trials per world.
a: 1,024-token secret; the probe's own time comes from probes followed by 16 listeners of 256 tokens, as in Figure~\ref{fig:copy}a, which favors that reading over placing it inside a ruler.
b: 4,096-token secret, both readings from the same trials.}
\label{fig:load}
\end{figure}

In the second and third runs, without load and up to half of capacity, the ruler reads secrets of 256, 1,024 and 4,096 tokens with advantage 0.980 to 1.000; Figure~\ref{fig:load} shows the latter two.
The probe's own time to first token, which hiding the hit request protects, is already weak there.
For a 1,024-token secret it gives $-0.03$ to 0.19 over the three runs at a quarter and half of capacity; in the second run its median differs by only 1 to 16\,ms between the worlds, against a standard deviation of 43 to 91\,ms.
It stays at 0.79 to 0.97 only for 4,096-token secrets.
The listeners thus read what the probe's own timing no longer shows.
Among trials in which every condition we could check from the logs held, 1,919 of 1,931 in the second and third runs were decided correctly; the checks approximate entry into the service by the clients' send times.
In an analysis made after the runs, an attacker that answers ``cached'' only when the rule decides~$H$ does so in 732 of 733 hit-world trials and in 1 of 782 miss-world trials up to half of capacity, a false-positive rate below 0.61\% at 95\% confidence.
By these bounds, a positive answer is right with probability at least 0.62 when one candidate in 100 is cached, and at least 0.14 when one in 1,000 is.

At 0.75 of capacity the advantage is 0.66 to 0.98 across the three runs, lowest where the trace segment keeps the cache fullest rather than where its request rate is highest.
There the probe missed because the victim's blocks had been evicted, in 2\% to 11\% of hit-world trials at a median cache use of 51\% to 57\% and in 26\% to 32\% at 94\%; the ruler's own 12,000 tokens before the probe add to the memory pressure.
In an exploratory analysis made after seeing the results, an attacker that fires only when its previous long request completed quickly would reach 0.98 to 1.00 at that load.
At 0.9 of capacity and above, honest traffic and the attacker's together overload the engine.

The results carry to other traffic and to a mixture-of-experts model.
Replaying instead the six busiest hours of the BurstGPT trace~\cite{wang2025burstgpt}, mostly short API calls, at 0.86 to 5.84 honest requests per second, the ruler reads secrets of 256 to 4,096 tokens with advantage 0.980 to 1.000, and none of the 1,119 trials in which every checked condition held was decided wrongly.
On the mixture-of-experts model DeepSeek-V2-Lite~\cite{deepseek2024v2}, whose step durations depend on content, it reads them with 0.948 to 1.000 up to half of capacity, 4.39 honest requests per second, where the probe's own first token gives at most 0.06 for a 1,024-token secret.

The ruler can take the shape of honest traffic, at some cost.
For 4,096-token secrets, taking output lengths from the trace leaves the advantage at 1.000, and taking input lengths from the trace gives 0.88 to 0.92, the loss coming from the coarser reading rather than from other requests.
Spacing every request by a nominal 5\,ms, about 170 per second as measured, gives 0.94 to 0.99 for 4,096-token secrets but 0.56 to 0.66 for 1,024; at 20\,ms apart, 4,096-token secrets give 0.69 to 0.84, because the longer window admits other requests.
When these runs are split after the fact by whether any honest request was sent in the window, all 284 trials without one whose listeners had equal lengths were decided correctly, a one-sided 95\% lower bound of 0.98 per world.

\begin{figure}[t]
\centering
\includegraphics{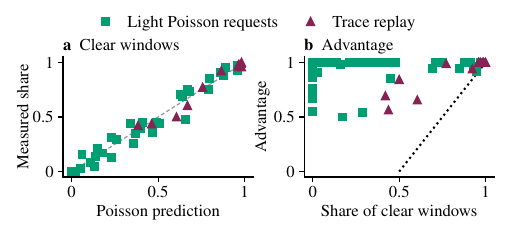}
\caption{Proposition~\ref{prop:burst} predicts how often the window stays clear, and the advantage stays near or above the bound this implies.
One point per setting: 49 with light Poisson requests of at most 1,024 prompt tokens at 2 to 100 per second on an RTX PRO 6000 GPU, secrets of 4,096 and 16,384 tokens and rulers sent at once or spaced by 3 to 30\,ms, each with at least 5 trials per world; 12 with the Azure trace at a quarter and half of capacity on an L40S GPU and equal-length listeners.
a: share of trials with no honest request in the window, against $e^{-\lambda w}$ for the measured rate $\lambda$ and window $w$; dashed: equality.
b: one-shot advantage against that share~$q$; dotted: $2q - 1$.}
\label{fig:window}
\end{figure}

Figure~\ref{fig:window} checks the first window condition against Proposition~\ref{prop:burst}: in 58 of its 61 settings, the share of trials whose window held no honest request is within 0.1 of $e^{-\lambda w}$.
The advantage never falls more than 0.05 below $2q - 1$, where $q$ is that share, the bound of Section~\ref{sec:needs} when the other conditions hold; settings with few trials can fall short of it by chance.
The bound is loose because a request that enters the window rarely makes the reading wrong: with light requests, settings in which at most a fifth of windows were clear still gave 0.50 to 1.00.
In absolute rates, rulers sent at once among light requests read 4,096-token secrets with advantage 1.00 at every rate up to 80 per second, and 16,384-token secrets with 0.91 to 1.00 up to 30 per second and 0.50 to 0.86 above, over 6 to 25 trials per world, although from 40 per second on most windows held an honest request.
A request that enters between the groups can make the reading wrong only if its length falls in a band set by the hit and the listener length, for a 1,024-token secret 496 to 1,520 tokens and its repeats about a step apart.
In analyses made after the runs with the trace's lengths, on one and two instances, 24 of 44 such interruptions inside the band were misread, against 3 of 180 outside it.

Clients on another node or behind a tunnel see the responses of one step spread by 0.1\,ms in the median and at most 0.7\,ms at the 99th percentile, and injecting that spread into the records leaves the reading unchanged.
Behind a TLS reverse proxy, with the attacker and the honest traffic on another node and responses streamed or buffered whole, the spread grows to about 1\,ms.
The ruler then reads 1,024- and 4,096-token secrets with 0.882 to 0.975 up to half of capacity, but at the operating point above its false-positive rate rises to 2.5\% to 2.9\%, from 0.13\% without the proxy; the spread stays far below the timestamp jitter the reading tolerates, and we have not isolated the cause.
On SGLang~0.5.16 with its default policy and 4,096-token chunks, the ruler reads secrets of 1,024 to 8,192 tokens with advantage 0.96 to 1.00 when idle and at a quarter of capacity.
At half of vLLM's capacity, which is closer to SGLang's own, its memory is nearly full and the probe hits in only 3\% to 17\% of hit-world trials, so the hit fails rather than the reading.

\subsection{Deployments}\label{sec:deploy}

We place two or four vLLM instances, each on its own L40S GPU with its own cache, behind the router of the vLLM production stack, and route the trace at the per-instance loads above.
Under its prefix-aware routing~\cite{vllm2026prefix}, a cache-affinity policy that sends a request to the instance holding its longest cached prefix, the probe goes where the victim's prefix is.
The attacker first sends prefixes that only it uses, to learn which of its requests share an instance, then runs one ruler per instance.
As Table~\ref{tab:deploy} shows, 4,096-token secrets are read with 0.90 to 0.98 on two and four instances, and 1,024-token secrets with 0.87 to 0.88 on two and 0.60 to 0.71 on four.

Routing that ignores the cache, here round robin, sends a single probe to the victim's instance only with the probability that they land together.
On two instances that probability is about one half, and the advantage falls by that factor; on four, round robin placed the probe with the victim in only 13\% of trials, and the advantage fell further.
This is not a defense that keeps the benefit: an honest request that reuses the secret is diluted by the same factor.

\begin{proposition}[Dilution]\label{prop:dilution}
Let an honest request carrying~$S$ be sent in the same state and under the same routing as the probe, and let $\pi$ be the probability that it hits~$S$ in~$H$.
Its expected saving is $\pi h$ tokens, and the lower bound $\Pr_H[G_H] + \Pr_M[G_M] - 1$ on a single probe's advantage under that routing carries the same factor~$\pi$, which is $1/N$ for uniform routing over $N$ instances.
Repeated reuse and repeated probes recover both, $i$ of them by the factor $1-(1-1/N)^i$.
\end{proposition}

An attacker that sends one copy of the probe to each instance recovers 0.34 to 0.84.
The attack carries to two other GPUs and to a large model, though not under identical conditions.
On RTX PRO 6000 GPUs, where vLLM's default budget is 8,192, rulers with four times longer listeners read 4,096- and 16,384-token secrets under cache affinity with advantage 0.75 to 0.93.
With the L40S listeners, also fixed before the run, 1,024-token secrets read only 0.14 to 0.45, and listeners lengthened for the larger budget would make the two worlds' phase readings overlap.
On one H100, where the default budget is also 8,192, rulers scaled to it read both lengths with 0.98 to 1.00 when idle and at a quarter of capacity, and 0.85 and 0.66 at half, where the engine runs near its memory limit and the probe hits in only 85\% and 77\% of hit-world trials; the 1,024-token ruler built for the L40S finds no step boundary on it.
Llama-3.1-70B in FP8 over two RTX PRO 6000 GPUs gives 0.98 to 1.00 when idle and at a quarter of capacity; at half, the attacker's own rulers of 60,000 to 80,000 tokens overload the engine and evict the victim's prefix before the probe in 37\% to 54\% of hit-world trials, so the advantage is 0.44 and 0.46.

\begin{table}[t]
\centering
\caption{One-shot advantage in other deployments}
\label{tab:deploy}
\footnotesize
\begin{tabular}{@{}llrr@{}}
\toprule
Setting & Secret & 0.25 & 0.5 \\
\midrule
2 instances, cache affinity & 1,024 & 0.88 & 0.87 \\
 & 4,096 & 0.98 & 0.94 \\
4 instances, cache affinity & 1,024 & 0.71 & 0.60 \\
 & 4,096 & 0.90 & 0.98 \\
2 instances, round robin, one probe & 1,024 & 0.45 & 0.45 \\
4 instances, round robin, one probe & 1,024 & 0.02 & 0.11 \\
2 instances, round robin, a probe each & 1,024 & 0.74 & 0.75 \\
 & 4,096 & 0.82 & 0.84 \\
4 instances, round robin, a probe each & 1,024 & 0.41 & 0.34 \\
 & 4,096 & 0.62 & 0.74 \\
RTX PRO 6000, 2 instances, cache affinity & 4,096 & 0.90 & 0.75 \\
 & 16,384 & 0.93 & 0.78 \\
70B, one instance on two GPUs & 4,096 & 0.98 & 0.44 \\
 & 16,384 & 1.00 & 0.46 \\
H100, one instance & 4,096 & 1.00 & 0.85 \\
 & 16,384 & 0.98 & 0.66 \\
\bottomrule
\end{tabular}
\tablenote{Columns 0.25 and 0.5 give the load as a fraction of capacity.
Secret lengths are in tokens; the first ten rows use L40S GPUs, the next four RTX PRO 6000 GPUs and the last two an H100 NVL; all rows use Llama-3.1-8B except the two 70B rows, which use Llama-3.1-70B.}
\end{table}

As an end-to-end case, a victim asks questions through the default retrieval-augmented pipeline of LlamaIndex~0.14.25~\cite{liu2022llamaindex} over 48 Wikipedia articles from SQuAD~\cite{rajpurkar2016squad}, split into 429 chunks, and the attacker asks which of four candidate chunks the victim retrieved first.
Allowed, in this case only, to know when the victim asks, it picks the right one in 97\%, 95\% and 85\% of cases when idle, at a quarter and at half of capacity, against 25\% by chance.
Membership reveals only that someone used the chunk recently: in an open-world estimate made afterwards, 1.9\% to 2.8\% of probes read one of the three wrong candidates as cached, in part because earlier probes had cached it.

\subsection{Limits}\label{sec:limits}

The shortest readable secret grows with the per-step budget.
The number of the ruler's requests grows as $B/h$, Section~\ref{sec:ruler}, and the sequence limit bounds how short listeners can be; on an L40S GPU, other listener lengths could take the theorem down to about 34 tokens but no lower, and on the RTX PRO 6000 a 1,024-token secret read at most 0.45.
Only on an idle engine does item~(iii) of Lemma~\ref{lem:copy} read single blocks.
Under routing that ignores the cache, one probe reads only the instance it lands on, at the cost in savings that Proposition~\ref{prop:dilution} states.
At high load the attack fails because the victim's blocks are evicted, not because the reading fails, and providers that guarantee retention remove this failure.
With several API servers in front of one engine, the attacker's own requests can enter out of order; at half of capacity, spacing the groups by 30\,ms restores the reading, and Appendix~\ref{app:experiments} gives the numbers.
Request-side reordering over wide-area paths was not measured.
Finally, we attack membership only: under load the theorem covers no block-by-block reconstruction of a prompt, and nothing here covers token-by-token reconstruction.

\section{Defenses: Visible Benefit or Indistinguishability}\label{sec:defenses}

\subsection{Three Knobs}\label{sec:knobs}

A defense sets three knobs.
The scope decides which blocks may be hit across accounts, from all of them down to none, which is per-account salting and what Pang et al.\ reach by partitioning the cache by user~\cite{pang2024cache}.
The release decides when each response leaves the service: when it is ready, after the hit request is held to its miss-world time, after a random delay, or on a schedule that hits do not move.
The reaction decides what the provider does once it suspects a probe, such as detecting, auditing, rate limiting, degrading or flushing.
Table~\ref{tab:defenses} places every other defense we found on these knobs in four groups; the groups organize the literature, overlap, and do not cover every design, such as ones that reorder requests.

Definition~\ref{def:security} splits every mechanism in two.
Either, in some workload, the view of some accounts within the rate budget is distributed differently in the two worlds, so the mechanism leaks with positive advantage, and attacks show how much; or no such view differs in any workload, and the mechanism is $(\beta,0)$-secure.
A defense in the first branch is judged by the advantage it leaves, and one in the second by what it costs and what tenants still gain, which Section~\ref{sec:inside} bounds.
Several entries are suggestions in attack papers or settings of a simulation rather than systems, so for each group we evaluate the mechanism that the closest published design describes and add its strongest version, an oracle that need not be implementable; where nothing has been published, only the oracle.
We measure first on an idle engine, a workload that Definition~\ref{def:security} covers, and then under honest load.

\begin{table}[t]
\centering
\caption{Published defenses and what stays readable}
\label{tab:defenses}
\footnotesize
\setlength{\tabcolsep}{4pt}
\begin{tabular}{@{}>{\raggedright\arraybackslash}p{0.30\columnwidth}ccc>{\raggedright\arraybackslash}p{0.46\columnwidth}@{}}
\toprule
Defense & Sc & Rl & Rx & What stays readable \\
\midrule
\multicolumn{5}{@{}l}{\emph{Hide the hit request, Section~\ref{sec:hidehit}}} \\
\addlinespace[2pt]
Gu et al.~\cite{gu2025auditing} & & \checkmark & & Copied saving, 0.96 to 0.99, vLLM and SGLang \\
InputSnatch~\cite{zheng2026inputsnatch} & \checkmark & \checkmark & \checkmark & Same; per-user scope is salting \\
PromptPeek~\cite{wu2025know} & & \checkmark & \checkmark & Same \\
\addlinespace[5pt]
\multicolumn{5}{@{}l}{\emph{Delay every response, Section~\ref{sec:delay}}} \\
\addlinespace[2pt]
KVGov noise~\cite{addagada2026governing} & & \checkmark & & Listeners: 0.99 at $\sigma = 20$\,ms \\
SpliceLeak noise~\cite{sun2026agent} & & \checkmark & & Averages out over listeners \\
Fonseca et al.~\cite{fonseca2026mitigacao} & & \checkmark & & Listeners cross bin edges \\
\addlinespace[5pt]
\multicolumn{5}{@{}l}{\emph{Narrow sharing or react, Section~\ref{sec:narrow}}} \\
\addlinespace[2pt]
SafeKV~\cite{chu2025selective} & \checkmark & & \checkmark & Content judged insensitive \\
CachePrune~\cite{wu2026cacheprune} & \checkmark & & & Retrieved documents \\
KVGov scope~\cite{addagada2026governing} & \checkmark & & \checkmark & Shared templates \\
Early Bird~\cite{song2025early} & \checkmark & & & Long enough candidates \\
SpliceDefense~\cite{sun2026agent} & \checkmark & & & Public document blocks \\
PrivKV-Sched~\cite{zuo2026privkvsched} & \checkmark & & & Admitted hits \\
CacheSolidarity~\cite{pennas2026cachesolidarity} & \checkmark & & \checkmark & Hit 0.98; flags 0.97 to 0.98 \\
Random threshold~\cite{harnik2010side} & \checkmark & & & Content once shareable \\
\addlinespace[5pt]
\multicolumn{5}{@{}l}{\emph{Keep the saving inside, Section~\ref{sec:inside}}} \\
\addlinespace[2pt]
Predictive mitigation~\cite{askarov2010predictive,zhang2011predictive} & & \checkmark & & Mispredictions \\
Deterministic execution~\cite{aviram2010determinating} & & \checkmark & & Imprecise timing, 0.19 to 0.98 in a microbenchmark \\
StopWatch~\cite{li2014stopwatch} & & \checkmark & & Nothing; replicas save nothing \\
VUsion~\cite{oliverio2017secure} & \checkmark & \checkmark & & Nothing; tenants gain nothing \\
\bottomrule
\end{tabular}
\tablenote{Sc, Rl and Rx mark the knobs a defense sets: scope, which blocks may be hit across accounts; release time of responses; and reaction after a suspected probe.
Advantages are one-shot, on vLLM unless stated.}
\end{table}

\subsection{Hiding the Hit Request}\label{sec:hidehit}

Gu et al.\ propose delaying hits so that they look like misses, noting that the provider still benefits because cached prompts take less GPU time~\cite{gu2025auditing}; InputSnatch suggests making response times uniform~\cite{zheng2026inputsnatch}, and PromptPeek keeping KV-cache activity from clients~\cite{wu2025know}.
None comes with an implementation, so we evaluate the oracle, which delaying hits alone may not even reach, Appendix~\ref{app:experiments}: everything about the probe is distributed as in the miss world, and scheduling is unchanged.
The GPU time the provider still saves is what the listeners read.
If every other response is released when ready, or at a time that does not depend on the world but falls before its miss-world readiness with positive probability, Lemma~\ref{lem:copy} gives the listeners positive advantage; Appendix~\ref{app:proofs} proves this.
The attacks of Section~\ref{sec:attack} never use the probe's response, so their results, Table~\ref{tab:deploy} included, are this oracle's evaluation.
A release time that never comes before the miss-world readiness instead keeps the saving inside, Section~\ref{sec:inside}.

\subsection{Delaying Every Response}\label{sec:delay}

Random delays target the gap between a hit and a miss.
KVGov adds Gaussian noise with a standard deviation of 20\,ms and reports that it stops a timing model of PromptPeek in its simulation, where a hit takes 5\,ms and a miss 50\,ms~\cite{addagada2026governing}; InputSnatch suggests injecting random delays~\cite{zheng2026inputsnatch}, and SpliceLeak considers random padding tokens while noting that such noise can be averaged out~\cite{sun2026agent}.
Classically, fuzzy time lowers the bandwidth of timing channels without closing them~\cite{hu1991reducing}, and random padding against BREACH averages out over repeated requests~\cite{gluck2013breach}.
The noise can be added to every response; the trouble is its scale.
It is set by the gap between one hit and one miss, whereas the listeners read a shift that alignment stretches to a whole step and that every listener carries.
Turning off streaming does not help either, since each listener returns one token.

\begin{theorem}[Price of random delays]\label{thm:delay}
Let every response be delayed independently of the world and of other responses, and let the attacker place $k$ listeners in the step before the probe's and $k$ in the step after it, so that the second group is ready later by the shift~$s$ in~$M$ than in~$H$, relative to the first.
(i) For Gaussian delays with standard deviation~$\sigma$ and Gaussian engine noise with standard deviation~$\sigma_e$ on the duration of the two steps, comparing the groups' mean release times gives advantage at least $2\Phi\bigl(s/(2\sqrt{\sigma_e^2 + 2\sigma^2/k})\bigr) - 1$.
(ii) For exponential delays with mean~$\mu$ and negligible engine noise, comparing the groups' earliest release times gives advantage at least $1 - e^{-ks/(2\mu)}$.
Keeping the advantage at most~$\varepsilon_0$ therefore requires $\sigma \ge \sqrt{k/2}\,\sqrt{s^2/(4z^2) - \sigma_e^2}$ with $z = \Phi^{-1}\bigl((1+\varepsilon_0)/2\bigr)$, or $\mu \ge ks/\bigl(2\ln(1/(1-\varepsilon_0))\bigr)$.
\end{theorem}

\begin{figure}[t]
\centering
\includegraphics{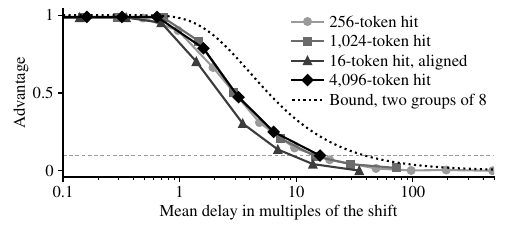}
\caption{Random delays must reach many times the shift.
Idle vLLM on an L40S GPU; each response gets an independent exponential delay, whose mean is given in multiples of the listeners' mean shift without delays, 10, 69, 144 and 312\,ms for the series in legend order.
Gray, light to dark: one-shot advantage of the mean first-token time of 16 listeners behind a 256-, 1,024- or 4,096-token hit, or of 8 listeners aligned to the budget behind a 16-token hit, cross-fitted over 140 to 172 trials per world and averaged over 20 draws of the delays.
Dotted: the lower bound of Theorem~\ref{thm:delay} for an attacker that compares the earliest releases of two groups of 8 listeners, a different statistic from the gray curves.
Dashed: advantage 0.1.}
\label{fig:delay}
\end{figure}

Appendix~\ref{app:proofs} gives the proof.
The rate budget limits~$k$: both groups and the probe enter in one burst, so $2k$ is at most the burst size, and the bounds grow with its square root and linearly in it.
With 16 requests per burst and $\varepsilon_0 = 0.1$, they ask for $\sigma$ of about 8 shifts and $\mu$ of about 38 shifts, 0.08 to 2.5\,s and 0.39 to 11.8\,s for the mean shifts of 10 to 312\,ms in Figure~\ref{fig:delay}.
By~(i), under KVGov's 20\,ms, one listener on each side of a full-step shift, as in the aligned construction of Section~\ref{sec:attack}, already guarantees advantage 0.994.
The theorem covers independent delays; for stationary Gaussian delays correlated across responses, the required deviation still grows with the logarithm of the number of listeners, Appendix~\ref{app:proofs}, and other correlated processes, such as one that shifts all responses after some time by a common amount, remain open.

Figure~\ref{fig:delay} adds independent exponential delays to the idle-engine records of Section~\ref{sec:attack}.
Measured in multiples of the shift, the four constructions fall together and reach advantage 0.1 at 9 to 16 times the shift, mean delays of 0.15 to 4.9\,s per response, with an attacker that averages its listeners; against the attacker of Theorem~\ref{thm:delay}, which compares earliest releases, the delays must reach 38 times.
Under KVGov's Gaussian noise the listeners keep 0.986 to 0.989 for hits of 1,024 tokens or more and in the aligned construction, where 20\,ms is 0.06 to 0.3 of the shift; only an unaligned 256-token hit, whose shift of 10\,ms is half the noise's deviation, falls to 0.67.
Under honest load the same noise added to the second run's recorded rulers lowers the ruler, which groups responses by the gaps between their arrivals, to 0.41 to 0.95, but assigning them to steps by the period between steps, which the same requests reveal, keeps 0.987 to 1.000 up to half of capacity.
Quantized release is no different: Fonseca et al.\ hold every response until the upper edge of a bin of time to first token~\cite{fonseca2026mitigacao}, and an attacker that knows the step boundaries can send a listener so that its first token falls just past a bin edge in~$M$, which the shift moves a bin earlier in~$H$.
Pacing admissions instead, one request per interval over all accounts, spreads the ruler as the attacker's own spacing did in Section~\ref{sec:load}, which approximates a pacer that honest requests seldom meet.
Under the trace, 5\,ms leaves 0.56 to 0.66 for 1,024-token secrets and 20\,ms leaves 0.69 to 0.84 for 4,096-token ones.
On an idle engine, however, no other request enters the longer window, and rulers spaced by up to 30\,ms read 1.00 on an RTX PRO 6000.
Pacing each account separately does not help either, since the attacker can spread its requests over more accounts.

\subsection{Narrowing Sharing and Reacting}\label{sec:narrow}

The other two knobs leave timing as it is.
Whether the membership of shared content can be read depends only on the release, so the ruler reads it as in Section~\ref{sec:attack}; the scope decides only which content that is.
Reactions come after the attacker has decided from one probe, and those that act per account are met by spreading its requests over a few accounts, Sections~\ref{sec:security} and~\ref{sec:needs}.
None of these defenses therefore gives a guarantee for membership, and we name what stays readable.

SafeKV keeps new blocks private until a detector clears them~\cite{chu2025selective}, and CachePrune shares only KV judged insensitive, at its strictest only fixed system prompts and retrieved documents~\cite{wu2026cacheprune}; content judged insensitive is read as before, and which document is retrieved depends on the private question, which the end-to-end case of Section~\ref{sec:deploy} reads.
KVGov can salt from the point where a request leaves a shared template~\cite{addagada2026governing}, so template membership stays readable.
Early Bird shares only prefixes above a minimum length~\cite{song2025early}, while a probe tests a whole candidate; SpliceDefense computes boundary blocks at worst case but shares public document blocks~\cite{sun2026agent}.
The authors of PrivKV-Sched, which budgets cross-tenant reuse, note that admitted hits may still reveal a prefix~\cite{zuo2026privkvsched}.
Random thresholds, carried over from deduplication, share a text only once a secret random number of accounts have written it~\cite{harnik2010side}, which leaves the attacker's own probe an advantage of about one over the number of possible thresholds.
But an attacker that writes the candidate from as many of its accounts as the largest threshold makes it shared, and where it also knows when the victim sends, as in the end-to-end case of Section~\ref{sec:deploy}, a workload that Definition~\ref{def:security} covers, the listeners read the victim's own hit.

CacheSolidarity records an owner for every block, serves the first cross-tenant hit as usual and then flags the block so that later hits on it are cut short~\cite{pennas2026cachesolidarity}.
The activator that switches it on is absent from the public code, so we keep protection always on, as the authors' security evaluation does and the setting least favorable to the attacker, with the authors' vLLM~0.8.5 on an L40S GPU.
The ruler reads the first cross-tenant hit with advantage 0.979 to 0.983.
The flags are shared state in the sense of Table~\ref{tab:exits}: where the victim's request leaves a flag decides how much of a later probe hits, which the attacker reads directly with 0.973, through aligned listeners with 0.980, and behind a decoy that turns it into 2,048 recomputed tokens with 0.981.
The public artifact does not clear flags when blocks are reused, so stale flags cut unrelated hits at random; Appendix~\ref{app:experiments} reports this defect, which costs sharing itself, and the one-line fix we applied.

The oracles of these knobs change nothing.
A perfect sensitivity classifier moves the guarantee onto the classifier, and a detector that knows the whole honest distribution cannot stop an attacker that imitates a class of rate-limited tenants, Section~\ref{sec:needs}; a classifier that refuses a request because it would hit makes the refusal an observation that differs between the worlds.

\subsection{Keeping the Saving Inside}\label{sec:inside}

The results of Sections~\ref{sec:hidehit} to~\ref{sec:narrow} share one cause.
A tenant receives from the service only the observations listed in Section~\ref{sec:security}, so any benefit it draws from shared content reaches it through them and is also a signal.

\begin{theorem}[Benefit is a signal]\label{thm:benefit}
Let account~$b$ be within the rate budget~$\beta$, and let the $V$ other accounts that share content with another account meet the victim conditions.
(i) If the mechanism is $(\beta,\varepsilon)$-secure, the total variation distance between $b$'s views in the workload and in the same workload with the shared texts of these $V$ accounts replaced by fresh ones, which has no cross-account sharing, is at most $V\varepsilon$.
(ii) If its one-shot advantage is at most~$\varepsilon$, replacing a cacheable text~$x$ by fresh text in the requests of the $k$ other accounts that send or hold it changes $b$'s view by at most $k\varepsilon$ in total variation.
Hence any utility of $b$'s view with values in an interval of length~$R$, such as its mean latency capped at a timeout, changes in expectation by at most $RV\varepsilon$, or by $Rk\varepsilon$ for~$x$; at $\varepsilon = 0$, $b$ gains nothing from content that other accounts wrote.
\end{theorem}

The proof, in Appendix~\ref{app:proofs}, chains Definition~\ref{def:security} over neighboring workloads, one account or one text at a time, with~$b$ itself as the observer and every other account sending its real traffic, so it needs neither an attacker that controls other traffic nor any assumption about how the provider uses identities.
It covers tenants within the rate budget and content of accounts that meet the victim conditions; whether a provider can leave the benefit to tenants above the rate budget while hiding it from everyone else is open.

Classical timing-channel defenses keep the saving inside, and Table~\ref{tab:defenses} lists the closest; we found none for LLM serving.
Predictive mitigation releases each event on a schedule that depends only on public inputs and past mispredictions, and widens it after each misprediction~\cite{askarov2010predictive,zhang2011predictive}.
Its bound counts bits leaked over a run, which says nothing about one bit; for one probe, the advantage is at most the sum over the two worlds of the probability of a misprediction, which also bounds how much earlier a hit can be released, Appendix~\ref{app:proofs}.
Replayed on the records of Section~\ref{sec:attack} with the strongest predictor, one that knows the attacker's construction, at its most conservative setting its bound keeps 1,024- and 4,096-token secrets below 0.1, at 95\% upper confidence limits of 0.074 and 0.028.
Responses in the hit world then arrive 10 to 12\,ms later on average than under salting.
For the aligned construction with hits of 16 and 256 tokens, its bound is 0.17 and 0.32, at 167 and 298\,ms later; under honest load it reaches 0.1 in only 2 of 8 pairs of load and construction, at 554 and 564\,ms later.
Deterministic execution releases results on a fixed period so that only public inputs set their timing~\cite{aviram2010determinating}; for an engine, a timed step releases each step's outputs at a deadline taken from a table of step durations plus a margin.
In a microbenchmark that runs the matrix multiplications of Llama-3.1-8B for full steps on an L40S GPU, margins of 0 and 2\% leave advantage 0.98 and 0.55: steps run faster after the idle time a hit leaves, so they overrun their deadlines less often in~$H$.
A margin of 5\% was unreadable, making every step 2.3\% longer than back-to-back execution.
A repetition without late releases still read 0.19 to 0.57 at nominal margins of 3\%, 5\% and 7\%, though nothing at 4\%; its calibration differed, so its 4\% deadline is the one closest to the first microbenchmark's 5\%, Appendix~\ref{app:experiments}; this reading comes from release-timer overshoot that varies with idle time by about 10\,$\mu$s on average.
Releasing at completion rather than at the deadline leaks 0.98, since the five steps after the idle time run about 45\,ms faster.
StopWatch releases at the median time of several replicas~\cite{li2014stopwatch}; replicas that actually run save nothing, and replicas that are only simulated are timed steps.
VUsion makes shared and unshared memory behave alike so that only the system keeps the saving~\cite{oliverio2017secure}, and Theorem~\ref{thm:benefit} says that at zero leakage the saving reaches no tenant within the rate budget; if the memory that sharing frees reaches tenants as longer retention of their own blocks, eviction order is an exit, Table~\ref{tab:exits}.

\begin{table*}[t]
\centering
\caption{The closest attacks and ours}
\label{tab:attacks}
\footnotesize
\begin{tabular}{@{}>{\raggedright\arraybackslash}p{0.11\textwidth}>{\raggedright\arraybackslash}p{0.22\textwidth}>{\raggedright\arraybackslash}p{0.18\textwidth}>{\raggedright\arraybackslash}p{0.23\textwidth}>{\raggedright\arraybackslash}p{0.16\textwidth}@{}}
\toprule
Attack & What it reads & Scheduling needed & Attacker's requests & Question \\
\midrule
PromptPeek~\cite{wu2025know} & Return order of the probe among dummy requests & Longest prefix match & Dummy batches around the probe, each over the batch limit & Token-by-token reconstruction \\
OptiLeak~\cite{wang2026optileak} & The same; under FCFS, the probe's time to first token & Longest prefix match & 20 dummies on each side and 20 candidates & Token-by-token reconstruction \\
SpliceLeak~\cite{sun2026agent} & Step-wave pattern in the probe's time to first token & Continuous batching with non-prefix reuse & As few as 63 requests per token & Length, then content \\
Early Bird~\cite{song2025early} & The probe's time to first token & None specific & 112.72 queries per token, with a fine-tuned predictor & System prompts, semantic caches \\
InputSnatch~\cite{zheng2026inputsnatch} & The probe's prefill time & None specific & 300 to 600 profiling runs, then probes per field & Field-by-field reconstruction \\
Gu et al.~\cite{gu2025auditing} & Distribution of the probe's time to first token & Black-box API & 250 samples per test & Whether and how widely a cache is shared \\
Wang et al.~\cite{wang2026rethinking} & Inter-token times of the attacker's own probes & FCFS with preemption and a global KV pool & Periodic probes & KV-cache usage, to time a denial of service \\
This paper & Steps in which its one-token listeners finish & FCFS steps under a token budget & A short burst of listeners before and after the probe & One-shot membership \\
\bottomrule
\end{tabular}
\tablenote{FCFS: admission first come, first served.
Request counts are as reported in each paper.}
\end{table*}

For LLM serving, zero leakage is attainable, so the second branch of Section~\ref{sec:knobs} is not empty.

\begin{proposition}[Lockstep replica]\label{prop:lockstep}
On one instance, let a replica run the engine's scheduler on the same requests while treating every cross-tenant hit as a miss, and let each replica step last $D$ of its shape, a bound on step durations fixed in advance.
Let the engine run each of the replica's batches without the tokens it can take from other accounts' cached blocks, wait until the replica's step ends, keep every block the replica holds, and release every output, admission, rejection, usage record and bill at the replica's times.
If every real step, with its output processing, ends within~$D$ of the replica step, each request's output tokens depend only on its own tokens and seed, errors are handled at the replica's times, releases are exact and the victim conditions hold, the mechanism is $(\beta,0)$-secure for every~$\beta$, and every tenant's view is distributed as in the workload without cross-account sharing.
\end{proposition}

The replica reads lengths, parameters, arrival times and self-hits but never cross-tenant matches, so its schedule does not depend on the world, and the engine, which runs a subset of each replica step, never falls behind it; Appendix~\ref{app:proofs} gives the proof.
This is the lock-step execution of secure multi-execution and StopWatch~\cite{devriese2010noninterference,li2014stopwatch} applied to one engine; output tokens that depend only on their own request need batch-invariant kernels~\cite{he2025defeating}, which vLLM~0.26 lacks, so the proposition assumes them, and we have not proved it for several instances.
The provider keeps the skipped work as idle GPU time, worth energy only if clocks are locked so that idle time does not speed up later steps, and pays the margin and release precision measured above.

This padding, the wait up to~$D$, cannot be avoided without knowing exactly how long skipped work would have taken.
Call a prefill block a potential hit if another account that meets the victim conditions could have written it earlier and it would still be cached, so that it would hit had that account sent the same prefix; a mechanism must hide potential hits whether or not they occur.
A busy period is a maximal interval in which the engine never idles, and prefill is saturated when every step fills its token budget.

\begin{theorem}[Price of padding]\label{thm:padding}
Consider mechanisms that schedule as the engine does under saturated prefill, skip cross-tenant hits without running work of the same shape from outside the request queues in their place, and only insert pauses or hold outputs, and let every account that shares content meet the victim conditions.
Suppose the provider knows each step shape's duration only within a width, independently across shapes.
If such a mechanism is $(\beta,0)$-secure under every duration model it cannot rule out, then under the fastest of them every output event of an account within the rate budget is released, in expectation, later than it would finish without sharing by at least the sum of the widths of the steps before it in its busy period that hold a potential hit whose shape has not run before, each step counted once, whether or not any hit occurs.
\end{theorem}

The proof turns each potential hit into a real one in a neighboring workload and uses Definition~\ref{def:security} to show that later releases cannot depend on that step's true duration.
Work of the same shape would remove the price, but a shape fixes where each sequence's computed tokens start, so in the usual case it means recomputing the hit request's own prefix, that is, not skipping.
A hit saves its prefill time once, 68\,ms for 1,024 tokens and about 300\,ms for 4,096 on an L40S GPU, while lockstep pauses grow by one width per step that holds a potential hit.
As an illustration, take a width of 0.13\% to 0.77\% of a full step, as for a table of step durations that we measured, which ignores the effects of idle time shown above, or of 8\% to 16\%, the held-out error of a law of time to first token that we fitted, and let saturated prefill hold a potential hit in every step.
The pauses then equal the saving of one 1,024-token hit within 0.4 to 50\,s and of one 4,096-token hit within 2 to 230\,s, and keep growing after that.
More hits spread the same pauses thinner, but at zero leakage the saving is idle time that only the provider can use, for energy or for its own work that never delays a release, whereas the pauses take wall-clock capacity.
A provider with an exact timing model pays less, as may one that learns shapes that recur, such as SGLang's fixed chunks, and reordering mechanisms lie outside the theorem.

\section{Discussion: Recommendations and Limits}\label{sec:discussion}

\subsection{Recommendations}\label{sec:recommend}

Providers that need isolation should salt per trust domain.
A provider knows only accounts, while the boundaries that matter may lie below them, between the end users of a relay or the teams and sessions of one customer; only the layer that knows such a boundary can draw it.
The interface should therefore accept an isolation key from that layer, and the engine should salt with it, as vLLM already does for a salt given per request~\cite{vllm2026security}.
A finer key forgoes more reuse, and by Theorem~\ref{thm:benefit} no way of hiding timing gives that reuse back at zero leakage.

Sharing across trust domains is then a choice to leak, not a timing problem to be engineered away: by Theorem~\ref{thm:benefit} a tenant's gain is bounded by the leakage allowed, so a provider that shares should offer it as an explicit option with that cost stated, rather than describe a defense as hiding hits.
A provider that wants only its own saving can run the lockstep replica of Proposition~\ref{prop:lockstep} and pay the padding of Theorem~\ref{thm:padding}, which suits deployments where energy matters more than capacity.

Defenses should be evaluated against what tenants observe, not against one request.
An evaluation should read every later request of accounts within the rate budget, report the one-shot advantage with one probe per secret, and let the attacker spread its requests over a few accounts.
Defenses that act only on release times can be evaluated without rerunning the engine, by replaying both worlds on records of an unmodified engine, as we did for random delays and predictive mitigation in Section~\ref{sec:defenses}.

\subsection{Limits}\label{sec:disc-limits}

Section~\ref{sec:limits} states the limits of the attack, Section~\ref{sec:defenses} the premises of our theorems, and Section~\ref{sec:threat} the models and victims we leave out.
Beyond these, we measured no deployed service, so its per-step budget, routing, batching and retention are unknown to us, and a larger budget lengthens the shortest readable secret.
Our attack reads the second exit of Table~\ref{tab:exits}; the other exits entered only through particular defenses in Section~\ref{sec:defenses}.
Honest load came from replaying one public trace on one to four instances, a second on one, and from synthetic Poisson requests.
Two further channels remain open.
In hybrid models mixing full and linear attention, under a cache shared across accounts, vLLM cuts a prefill at the end of a prefix that another account has cached, to store the linear state there, so a request that hits nothing still takes 11 to 25\,ms longer, Appendix~\ref{app:experiments}.
When instances share one KV pool, a hit can occur on any instance, and we have not measured the ruler there.

\section{Related Work}\label{sec:related}

Table~\ref{tab:attacks} compares the closest attacks with ours; the probe's own latency also reveals hits behind a shared KV layer whose keys drop the salt~\cite{song2026preserving}.

Like PromptPeek and OptiLeak~\cite{wu2025know,wang2026optileak}, our attacker surrounds the probe with its own requests and reads a discrete signal; there, longest-prefix-match scheduling moves a hit ahead of the dummies, whereas our listeners need only first-come-first-served steps and read the saving that reaches the requests behind the probe.
Wang et al.\ read global KV-cache usage from their own probes' inter-token times to time a denial-of-service attack, and bound one prefill's average delay to a co-located request by about its cost divided by the number of batch slots~\cite{wang2026rethinking}; requests already queued behind a prefill under a token budget instead shift by the whole saving.
Integrity attacks poison or hijack shared caches~\cite{wu2026cache,zhang2026hijackkv}; other side channels read packet sizes, the timing of speculative or parallel decoding, or hardware caches~\cite{weiss2024what,wei2024when,carlini2024remote,gao2025i}, and under expert-choice routing a white-box attacker that controls a victim's batch reads its prompt~\cite{yona2024stealing}.

Section~\ref{sec:defenses} places the published defenses, which aim to keep reuse while hiding hits; some let the attacker hold many colluding accounts~\cite{chu2025selective,pennas2026cachesolidarity}.
All evaluate the leak through the probe's own response; some mention interference or ordering among concurrent requests or define security over all latency the attacker observes~\cite{song2025early,sun2026agent,pang2024cache,zuo2026privkvsched}, but none considers where a hit's saving goes, and Appendix~\ref{app:defense-tm} quotes each threat model.

Reading another party's work through one's own requests in a shared queue is classical.
A spy that enqueues short tasks in a shared event loop sees whether and how long a victim ran between them~\cite{vila2017loophole}, and a ping sent right after a transaction reveals the extra work the transaction caused~\cite{tramer2020remote}.
Packets queued behind a victim's reveal its bursts under first-in-first-out queueing and, with denser probes, under any work-conserving scheduler~\cite{beams2021packet}, whose leakage has been quantified~\cite{kadloor2013delay,kadloor2016mitigating,gong2016quantifying}.
In these channels the signal is the victim's own activity or work caused by a secret.
Our listeners instead read a saving that another tenant's cached text gives the attacker's own probe, while that tenant may be offline, and step-based engines copy this saving to every queued request and round it to whole steps.

As with compression and deduplication, a saving the requester can observe is a signal~\cite{gluck2013breach,harnik2010side,bosman2016dedup,schwarzl2022remote}; in LLM serving it is observed as time, and also by requests other than the one that received it.
Privacy has been shown to cost the benefit in client-side deduplication~\cite{armknecht2017side}, in named-data caches~\cite{acs2013cache}, and under differential privacy for a cache's sequence of misses~\cite{ghazi2026on}; our bound instead measures the gain on what a tenant itself receives, against rate-limited attackers that hold several accounts, whatever the provider does with identities.
Keeping the saving inside does not stop leakage once it changes physical state: under isolated per-connection service, total load still heats a CPU and shifts its clock skew, read remotely from timestamps~\cite{murdoch2006hot}.

Predictive mitigation, deterministic execution, time protection and secure page fusion limit how timing depends on secrets or on shared pages~\cite{askarov2010predictive,zhang2011predictive,aviram2010determinating,ge2019time,oliverio2017secure}; applied to caching, they keep the saving inside the system.
Our zero-leakage construction follows secure multi-execution, which lets a copy that never sees the secret produce the public outputs~\cite{devriese2010noninterference}, and StopWatch, which makes an attacker's timings match a replica not co-resident with the victim~\cite{li2013mitigating,li2014stopwatch}.
Our replica only runs the scheduler, without cross-tenant hits, and fixes batches, cached blocks and release times; because skipped work has only approximately known duration, its padding is paid per step, not per hit, as in constant-rate traffic padding~\cite{dyer2012peek}.

Requiring identically distributed later views is noninterference~\cite{goguen1982security} applied to cache hits; our prior work KeyPooling holds relays to the same standard~\cite{sun2026keypooling}, and Beams et al.\ define scheduler privacy over the whole view~\cite{beams2021packet}.

\section{Conclusion}\label{sec:conclusion}

A cache hit saves work, and hiding the hit does not make the saving disappear.
Work-conserving step schedulers copy it to every request queued behind the hit, where listener requests of an attacker within ordinary rate limits read it with one probe.
Defenses that make the hit request look like a miss or narrow what is shared therefore give no guarantee against membership inference, and random delays cut the advantage to 0.1 only at about ten times the shift or more.
The cause is not particular to these defenses: whatever benefit a tenant observes from shared content is also a signal, so its gain from content written by $k$ other accounts is at most of order $k$ times the leakage allowed.
At zero leakage only the provider keeps the saving, and hiding it costs padding that accrues with time rather than with hits.
Providers that need isolation should salt per trust domain, with the boundary drawn by the layer that knows it.

\bibliographystyle{IEEEtran}
\bibliography{refs}

@inproceedings{acs2013cache,
  author = {Gergely Acs and Mauro Conti and Paolo Gasti and Cesar Ghali and Gene Tsudik},
  title = {Cache Privacy in Named-Data Networking},
  booktitle = {Proc. IEEE ICDCS},
  pages = {41--51},
  year = {2013},
  doi = {10.1109/ICDCS.2013.12}
}

@misc{addagada2026governing,
  author = {Tejasvi C. Addagada},
  title = {Governing the {KV} Cache: Preventing Timing Side-Channel Leakage in Multi-Tenant {LLM} Inference},
  year = {2026},
  eprint = {2608.09225},
  archivePrefix = {arXiv},
  primaryClass = {cs.CR},
  howpublished = {arXiv preprint arXiv:2608.09225},
  note = {Introduces KVGov}
}

@inproceedings{agrawal2024taming,
  author = {Amey Agrawal and Nitin Kedia and Ashish Panwar and Jayashree Mohan and Nipun Kwatra and Bhargav Gulavani and Alexey Tumanov and Ramachandran Ramjee},
  title = {Taming {Throughput-Latency} Tradeoff in {LLM} Inference with {Sarathi-Serve}},
  booktitle = {Proc. OSDI},
  pages = {117--134},
  year = {2024}
}

@misc{anthropic2026prompt,
  author = {{Anthropic}},
  title = {Prompt caching},
  howpublished = {Claude Platform documentation},
  year = {2026},
  url = {https://platform.claude.com/docs/en/build-with-claude/prompt-caching},
  note = {Section ``Cache storage and sharing''. Accessed: 2026-10-26}
}

@inproceedings{armknecht2017side,
  author = {Frederik Armknecht and Colin Boyd and Gareth T. Davies and Kristian Gj{\o}steen and Mohsen Toorani},
  title = {Side Channels in Deduplication: Trade-offs between Leakage and Efficiency},
  booktitle = {Proc. ACM AsiaCCS},
  pages = {266--274},
  month = apr,
  year = {2017},
  doi = {10.1145/3052973.3053019}
}

@inproceedings{askarov2010predictive,
  author = {Aslan Askarov and Danfeng Zhang and Andrew C. Myers},
  title = {Predictive Black-Box Mitigation of Timing Channels},
  booktitle = {Proc. ACM CCS},
  pages = {297--307},
  month = oct,
  year = {2010},
  doi = {10.1145/1866307.1866341}
}

@inproceedings{aviram2010determinating,
  author = {Amittai Aviram and Sen Hu and Bryan Ford and Ramakrishna Gummadi},
  title = {Determinating Timing Channels in Compute Clouds},
  booktitle = {Proc. ACM CCSW},
  pages = {103--108},
  month = oct,
  year = {2010},
  doi = {10.1145/1866835.1866854}
}

@inproceedings{beams2021packet,
  author = {Andrew Beams and Sampath Kannan and Sebastian Angel},
  title = {Packet Scheduling with Optional Client Privacy},
  booktitle = {Proc. ACM CCS},
  pages = {3415--3430},
  year = {2021},
  doi = {10.1145/3460120.3485371}
}

@inproceedings{bosman2016dedup,
  author = {Erik Bosman and Kaveh Razavi and Herbert Bos and Cristiano Giuffrida},
  title = {Dedup Est Machina: Memory Deduplication as an Advanced Exploitation Vector},
  booktitle = {Proc. IEEE S\&P},
  pages = {987--1004},
  month = may,
  year = {2016},
  doi = {10.1109/SP.2016.63}
}

@inproceedings{carlini2022membership,
  author = {Nicholas Carlini and Steve Chien and Milad Nasr and Shuang Song and Andreas Terzis and Florian Tram{\`e}r},
  title = {Membership Inference Attacks From First Principles},
  booktitle = {Proc. IEEE S\&P},
  pages = {1897--1914},
  year = {2022},
  doi = {10.1109/SP46214.2022.9833649}
}

@misc{carlini2024remote,
  author = {Nicholas Carlini and Milad Nasr},
  title = {Remote Timing Attacks on Efficient Language Model Inference},
  year = {2024},
  eprint = {2410.17175},
  archivePrefix = {arXiv},
  primaryClass = {cs.CR},
  howpublished = {arXiv preprint arXiv:2410.17175}
}

@misc{chu2025selective,
  author = {Kexin Chu and Zecheng Lin and Dawei Xiang and Zixu Shen and Jianchang Su and Cheng Chu and Yiwei Yang and Wenhui Zhang and Wenfei Wu and Wei Zhang},
  title = {Selective {KV}-Cache Sharing to Mitigate Timing Side-Channels in {LLM} Inference},
  year = {2025},
  eprint = {2508.08438},
  archivePrefix = {arXiv},
  primaryClass = {cs.CR},
  howpublished = {arXiv preprint arXiv:2508.08438},
  note = {Introduces SafeKV}
}

@misc{deepseek2024v2,
  author = {{DeepSeek-AI}},
  title = {{DeepSeek-V2}: A Strong, Economical, and Efficient Mixture-of-Experts Language Model},
  year = {2024},
  eprint = {2405.04434},
  archivePrefix = {arXiv},
  howpublished = {arXiv preprint arXiv:2405.04434}
}

@inproceedings{devriese2010noninterference,
  author = {Dominique Devriese and Frank Piessens},
  title = {Noninterference through Secure Multi-execution},
  booktitle = {Proc. IEEE S\&P},
  pages = {109--124},
  year = {2010},
  doi = {10.1109/SP.2010.15}
}

@inproceedings{dyer2012peek,
  author = {Kevin P. Dyer and Scott E. Coull and Thomas Ristenpart and Thomas Shrimpton},
  title = {Peek-a-Boo, {I} Still See You: Why Efficient Traffic Analysis Countermeasures Fail},
  booktitle = {Proc. IEEE S\&P},
  pages = {332--346},
  year = {2012},
  doi = {10.1109/SP.2012.28}
}

@misc{fahey2026cacheprobe,
  author = {Ryan Fahey},
  title = {{CacheProbe}: Auditing Prompt Cache Isolation in Gateway {API}s},
  year = {2026},
  eprint = {2605.30613},
  archivePrefix = {arXiv},
  primaryClass = {cs.CR},
  howpublished = {arXiv preprint arXiv:2605.30613},
  note = {Accepted at the SAGAI 2026 workshop}
}

@misc{fireworks2026prompt,
  author = {{Fireworks AI}},
  title = {Prompt caching},
  howpublished = {Fireworks AI documentation},
  year = {2026},
  url = {https://docs.fireworks.ai/guides/prompt-caching},
  note = {Section ``Data privacy''. Accessed: 2026-10-26}
}

@inproceedings{fonseca2026mitigacao,
  author = {Bruno Fonseca and Uelinton Brezolin and Michele Nogueira},
  title = {Mitiga{\c{c}}{\~a}o de Ataques de Canal Lateral Temporal em Sistemas de Serving de {LLMs} por Meio da Equaliza{\c{c}}{\~a}o de Lat{\^e}ncia},
  booktitle = {Proc. SBSeg},
  pages = {738--753},
  year = {2026},
  doi = {10.5753/sbseg.2026.29354},
  note = {In Portuguese, with an English abstract}
}

@inproceedings{gao2025i,
  author = {Zibo Gao and Junjie Hu and Feng Guo and Yixin Zhang and Yinglong Han and Siyuan Liu and Haiyang Li and Zhiqiang Lv},
  title = {{I} Know What You Said: Unveiling Hardware Cache Side-Channels in Local Large Language Model Inference},
  booktitle = {Proc. USENIX Security},
  pages = {1649--1668},
  month = aug,
  year = {2025},
  isbn = {978-1-939133-52-6}
}

@inproceedings{ge2019time,
  author = {Qian Ge and Yuval Yarom and Tom Chothia and Gernot Heiser},
  title = {Time Protection: The Missing {OS} Abstraction},
  booktitle = {Proc. EuroSys},
  pages = {1--17},
  month = mar,
  year = {2019},
  doi = {10.1145/3302424.3303976}
}

@misc{ghazi2026on,
  author = {Badih Ghazi and Ravi Kumar and Pasin Manurangsi and Jelani Nelson and Adam Sealfon and Samson Zhou},
  title = {On Differential Privacy and Caching},
  howpublished = {Theory and Practice of Differential Privacy Workshop (TPDP 2026)},
  year = {2026}
}

@misc{gluck2013breach,
  author = {Yoel Gluck and Neal Harris and Angelo Prado},
  title = {{BREACH}: Reviving the {CRIME} Attack},
  howpublished = {White paper; presented at Black Hat USA 2013 as ``SSL, gone in 30 seconds -- a BREACH beyond CRIME''},
  month = jul,
  year = {2013},
  url = {https://breachattack.com/},
  note = {Paper dated 2013-07-12. Accessed: 2026-10-26}
}

@inproceedings{goguen1982security,
  author = {J. A. Goguen and J. Meseguer},
  title = {Security Policies and Security Models},
  booktitle = {Proc. IEEE Symp. Security and Privacy},
  pages = {11--20},
  month = apr,
  year = {1982},
  doi = {10.1109/SP.1982.10014}
}

@article{gong2016quantifying,
  author = {Xun Gong and Negar Kiyavash},
  title = {Quantifying the Information Leakage in Timing Side Channels in Deterministic Work-Conserving Schedulers},
  journal = {IEEE/ACM Trans. Netw.},
  volume = {24},
  number = {3},
  pages = {1841--1852},
  month = jun,
  year = {2016},
  doi = {10.1109/TNET.2015.2438860}
}

@inproceedings{gu2025auditing,
  author = {Chenchen Gu and Xiang Lisa Li and Rohith Kuditipudi and Percy Liang and Tatsunori Hashimoto},
  title = {Auditing Prompt Caching in Language Model {API}s},
  booktitle = {Proc. ICML},
  volume = {267},
  pages = {20477--20496},
  month = jul,
  year = {2025}
}

@article{harnik2010side,
  author = {Danny Harnik and Benny Pinkas and Alexandra Shulman-Peleg},
  title = {Side Channels in Cloud Services: Deduplication in Cloud Storage},
  journal = {IEEE Security \& Privacy},
  volume = {8},
  number = {6},
  pages = {40--47},
  month = nov,
  year = {2010},
  doi = {10.1109/MSP.2010.187}
}

@misc{he2025defeating,
  author = {Horace He and {Thinking Machines Lab}},
  title = {Defeating Nondeterminism in {LLM} Inference},
  howpublished = {Thinking Machines Lab: Connectionism (blog)},
  month = sep,
  year = {2025},
  doi = {10.64434/tml.20250910},
  url = {https://thinkingmachines.ai/blog/defeating-nondeterminism-in-llm-inference/},
  note = {Accessed: 2026-10-26}
}

@inproceedings{hu1991reducing,
  author = {Wei-Ming Hu},
  title = {Reducing Timing Channels with Fuzzy Time},
  booktitle = {Proc. IEEE Symp. Research in Security and Privacy},
  pages = {8--20},
  year = {1991},
  doi = {10.1109/RISP.1991.130768}
}

@inproceedings{kadloor2013delay,
  author = {Sachin Kadloor and Negar Kiyavash},
  title = {Delay optimal policies offer very little privacy},
  booktitle = {Proc. IEEE INFOCOM},
  pages = {2454--2462},
  month = apr,
  year = {2013},
  doi = {10.1109/INFCOM.2013.6567051}
}

@article{kadloor2016mitigating,
  author = {Sachin Kadloor and Negar Kiyavash and Parv Venkitasubramaniam},
  title = {Mitigating Timing Side Channel in Shared Schedulers},
  journal = {IEEE/ACM Trans. Netw.},
  volume = {24},
  number = {3},
  pages = {1562--1573},
  month = jun,
  year = {2016},
  doi = {10.1109/TNET.2015.2418194}
}

@inproceedings{li2013mitigating,
  author = {Peng Li and Debin Gao and Michael K. Reiter},
  title = {Mitigating Access-Driven Timing Channels in Clouds using {StopWatch}},
  booktitle = {Proc. IEEE/IFIP DSN},
  pages = {1--12},
  month = jun,
  year = {2013},
  doi = {10.1109/DSN.2013.6575299}
}

@article{li2014stopwatch,
  author = {Peng Li and Debin Gao and Michael K. Reiter},
  title = {{StopWatch}: A Cloud Architecture for Timing Channel Mitigation},
  journal = {ACM Trans. Inf. Syst. Secur.},
  volume = {17},
  number = {2},
  articleno = {8},
  pages = {8:1--8:28},
  month = nov,
  year = {2014},
  doi = {10.1145/2670940}
}

@inproceedings{lin2026behavioral,
  author = {Guanjie Lin and Yinxin Wan and Shichao Pei and Ting Xu and Kuai Xu and Guoliang Xue},
  title = {Behavioral Consistency and Transparency Analysis on Large Language Model {API} Gateways},
  booktitle = {Proc. ACM IMC},
  pages = {563--573},
  year = {2026},
  doi = {10.1145/3777912.3809156}
}

@misc{liu2022llamaindex,
  author = {Jerry Liu},
  title = {{LlamaIndex}},
  howpublished = {Software},
  url = {https://github.com/run-llama/llama_index},
  year = {2022},
  note = {Version 0.14.25 used}
}

@inproceedings{murdoch2006hot,
  author = {Steven J. Murdoch},
  title = {Hot or Not: Revealing Hidden Services by their Clock Skew},
  booktitle = {Proc. ACM CCS},
  pages = {27--36},
  year = {2006},
  doi = {10.1145/1180405.1180410}
}

@inproceedings{oliverio2017secure,
  author = {Marco Oliverio and Kaveh Razavi and Herbert Bos and Cristiano Giuffrida},
  title = {Secure Page Fusion with {VUsion}},
  booktitle = {Proc. SOSP},
  pages = {531--545},
  month = oct,
  year = {2017},
  doi = {10.1145/3132747.3132781}
}

@misc{openai2026prompt,
  author = {{OpenAI}},
  title = {Prompt caching},
  howpublished = {OpenAI API documentation},
  year = {2026},
  url = {https://developers.openai.com/api/docs/guides/prompt-caching},
  note = {Section ``Cache location''. Accessed: 2026-10-26}
}

@misc{openai2026rate,
  author = {{OpenAI}},
  title = {Best practices for managing rate limits in the {API}},
  howpublished = {OpenAI Help Center, article 6891753},
  url = {https://help.openai.com/en/articles/6891753},
  year = {2026},
  note = {Accessed: 2026-10-25}
}

@inproceedings{pang2024cache,
  author = {Zixuan Pang and Wenhao Wang and Yong Liao},
  title = {Cache Partitioning for Mitigating Timing Side-Channel Attacks in {LLM} Serving Systems},
  booktitle = {Proc. ICFTIC},
  pages = {1238--1245},
  year = {2024},
  doi = {10.1109/ICFTIC64248.2024.10913329}
}

@inproceedings{patel2024splitwise,
  author = {Pratyush Patel and Esha Choukse and Chaojie Zhang and Aashaka Shah and {\'I}{\~n}igo Goiri and Saeed Maleki and Ricardo Bianchini},
  title = {Splitwise: Efficient Generative {LLM} Inference Using Phase Splitting},
  booktitle = {Proc. ACM/IEEE ISCA},
  pages = {118--132},
  year = {2024},
  doi = {10.1109/ISCA59077.2024.00019}
}

@misc{pennas2026cachesolidarity,
  author = {Panagiotis Georgios Pennas and Konstantinos Papaioannou and Marco Guarnieri and Thaleia Dimitra Doudali},
  title = {{CacheSolidarity}: Preventing Prefix Caching Side Channels in Multi-Tenant {LLM} Serving Systems},
  year = {2026},
  eprint = {2603.10726},
  archivePrefix = {arXiv},
  primaryClass = {cs.CR},
  howpublished = {arXiv preprint arXiv:2603.10726},
  note = {To appear at ACM/IFIP Middleware 2026}
}

@inproceedings{rajpurkar2016squad,
  author = {Pranav Rajpurkar and Jian Zhang and Konstantin Lopyrev and Percy Liang},
  title = {{SQuAD}: 100,000+ Questions for Machine Comprehension of Text},
  booktitle = {Proc. EMNLP},
  pages = {2383--2392},
  year = {2016},
  doi = {10.18653/v1/D16-1264}
}

@inproceedings{schwarzl2022remote,
  author = {Martin Schwarzl and Erik Kraft and Moritz Lipp and Daniel Gruss},
  title = {Remote Memory-Deduplication Attacks},
  booktitle = {Proc. NDSS},
  month = apr,
  year = {2022},
  doi = {10.14722/ndss.2022.23081}
}

@article{song2025early,
  author = {Linke Song and Zixuan Pang and Wenhao Wang and Zihao Wang and XiaoFeng Wang and Hongbo Chen and Wei Song and Yier Jin and Dan Meng and Rui Hou},
  title = {The Early Bird Catches the Leak: Unveiling Timing Side Channels in {LLM} Serving Systems},
  journal = {IEEE Trans. Inf. Forensics Security},
  volume = {20},
  pages = {11431--11446},
  year = {2025},
  doi = {10.1109/TIFS.2025.3622954}
}

@misc{song2026preserving,
  author = {Wei Song and Yuxin Cao and Xi Zheng and Leo Zhang and Xiao Cheng},
  title = {Preserving Provenance in Shared {KV} Caches for {LLM} Serving},
  year = {2026},
  eprint = {2609.38706},
  archivePrefix = {arXiv},
  primaryClass = {cs.DC},
  howpublished = {arXiv preprint arXiv:2609.38706}
}

@misc{sun2026agent,
  author = {He Sun and Shinan Liu and Siyuan Ma and Junhao Li and Mingjun Xiao and Wenhao Jiang},
  title = {Agent-Assisted Side-Channel Attacks on Non-Prefix {KV} Cache in {RAG}},
  year = {2026},
  eprint = {2606.21842},
  archivePrefix = {arXiv},
  primaryClass = {cs.CR},
  howpublished = {arXiv preprint arXiv:2606.21842},
  note = {Introduces the SpliceLeak attack and the SpliceDefense mitigation}
}

@misc{sun2026keypooling,
  author = {Bowen Sun and Yixi Cai and Xiaogeng Liu and Zhengyue Zhao and Yinzhi Cao and Chaowei Xiao},
  title = {{KeyPooling}: Measuring Where {LLM} {API} Relay Paths Collapse Prompt Cache Isolation},
  year = {2026},
  eprint = {2608.17485},
  archivePrefix = {arXiv},
  primaryClass = {cs.CR},
  howpublished = {arXiv preprint arXiv:2608.17485}
}

@inproceedings{tramer2020remote,
  author = {Florian Tram{\`e}r and Dan Boneh and Kenneth G. Paterson},
  title = {Remote Side-Channel Attacks on Anonymous Transactions},
  booktitle = {Proc. USENIX Security},
  pages = {2739--2756},
  year = {2020}
}

@inproceedings{vila2017loophole,
  author = {Pepe Vila and Boris K{\"o}pf},
  title = {Loophole: Timing Attacks on Shared Event Loops in {Chrome}},
  booktitle = {Proc. USENIX Security},
  pages = {849--864},
  year = {2017}
}

@misc{vllm2026automatic,
  author = {{vLLM Project}},
  title = {Automatic Prefix Caching},
  howpublished = {vLLM documentation (latest), design document},
  year = {2026},
  url = {https://docs.vllm.ai/en/latest/design/prefix_caching/},
  note = {Paragraph ``Cache Isolation for Security'' in the introduction. Accessed: 2026-10-26}
}

@misc{vllm2026prefix,
  author = {{vLLM Production Stack}},
  title = {Prefix Aware Routing},
  howpublished = {vLLM Production Stack documentation},
  url = {https://docs.vllm.ai/projects/production-stack/en/latest/use_cases/prefix-aware-routing.html},
  year = {2026},
  note = {Router option routingLogic prefixaware. Accessed: 2026-10-29}
}

@misc{vllm2026security,
  author = {{vLLM Project}},
  title = {Security},
  howpublished = {vLLM documentation (latest), usage guide},
  year = {2026},
  url = {https://docs.vllm.ai/en/latest/usage/security/},
  note = {Sections ``Multi-Tenant Deployments'' and ``Prefix Cache Timing Side-Channel Mitigation (Cache Salting)''. Accessed: 2026-10-26}
}

@inproceedings{wang2025burstgpt,
  author = {Yuxin Wang and Yuhan Chen and Zeyu Li and Xueze Kang and Yuchu Fang and Yeju Zhou and Yang Zheng and Zhenheng Tang and Xin He and Rui Guo and Xin Wang and Qiang Wang and Amelie Chi Zhou and Xiaowen Chu},
  title = {{BurstGPT}: A Real-World Workload Dataset to Optimize {LLM} Serving Systems},
  booktitle = {Proc. ACM KDD},
  pages = {5831--5841},
  year = {2025},
  doi = {10.1145/3711896.3737413}
}

@misc{wang2026optileak,
  author = {Longxiang Wang and Xiang Zheng and Xuhao Zhang and Yao Zhang and Ye Wu and Cong Wang},
  title = {{OptiLeak}: Efficient Prompt Reconstruction via Reinforcement Learning in Multi-tenant {LLM} Services},
  year = {2026},
  eprint = {2602.20595},
  archivePrefix = {arXiv},
  primaryClass = {cs.CR},
  howpublished = {arXiv preprint arXiv:2602.20595}
}

@inproceedings{wang2026rethinking,
  author = {Tianyi Wang and Huawei Fan and Yuanchao Shu and Peng Cheng and Cong Wang},
  title = {Rethinking Latency Denial-of-Service: Attack the {LLM} Serving Framework, Not the Model},
  booktitle = {Proc. NeurIPS},
  year = {2026},
  note = {To appear}
}

@misc{wei2024when,
  author = {Jiankun Wei and Abdulrahman Abdulrazzag and Tianchen Zhang and Adel Muursepp and Gururaj Saileshwar},
  title = {When Speculation Spills Secrets: Side Channels via Speculative Decoding In {LLMs}},
  year = {2024},
  eprint = {2411.01076},
  archivePrefix = {arXiv},
  primaryClass = {cs.CL},
  howpublished = {arXiv preprint arXiv:2411.01076}
}

@inproceedings{weiss2024what,
  author = {Roy Weiss and Daniel Ayzenshteyn and Guy Amit and Yisroel Mirsky},
  title = {What Was Your Prompt? {A} Remote Keylogging Attack on {AI} Assistants},
  booktitle = {Proc. USENIX Security},
  pages = {3367--3384},
  month = aug,
  year = {2024},
  isbn = {978-1-939133-44-1}
}

@inproceedings{wu2025know,
  author = {Guanlong Wu and Zheng Zhang and Yao Zhang and Weili Wang and Jianyu Niu and Ye Wu and Yinqian Zhang},
  title = {I Know What You Asked: Prompt Leakage via {KV}-Cache Sharing in Multi-Tenant {LLM} Serving},
  booktitle = {Proc. NDSS},
  month = feb,
  year = {2025},
  doi = {10.14722/ndss.2025.241772}
}

@inproceedings{wu2026cache,
  author = {XiangFan Wu and Lingyun Ying and Guoqiang Chen and Yacong Gu and Haipeng Qu},
  title = {Cache Me, Catch You: Cache Related Security Threats in {LLM} Serving Frameworks},
  booktitle = {Proc. NDSS},
  month = feb,
  year = {2026},
  isbn = {979-8-9919276-8-0},
  doi = {10.14722/ndss.2026.242812}
}

@misc{wu2026cacheprune,
  author = {Guanlong Wu and Zhaohan Li and Yao Zhang and Zheng Zhang and Jianyu Niu and Ye Wu and Yinqian Zhang},
  title = {{CachePrune}: Privacy-Aware and Fine-Grained {KV} Cache Sharing for Efficient {LLM} Inference},
  year = {2026},
  eprint = {2605.23640},
  archivePrefix = {arXiv},
  primaryClass = {cs.CR},
  howpublished = {arXiv preprint arXiv:2605.23640}
}

@inproceedings{yeom2018privacy,
  author = {Samuel Yeom and Irene Giacomelli and Matt Fredrikson and Somesh Jha},
  title = {Privacy Risk in Machine Learning: Analyzing the Connection to Overfitting},
  booktitle = {Proc. IEEE CSF},
  pages = {268--282},
  year = {2018},
  doi = {10.1109/CSF.2018.00027}
}

@misc{yona2024stealing,
  author = {Itay Yona and Ilia Shumailov and Jamie Hayes and Nicholas Carlini},
  title = {Stealing User Prompts from Mixture of Experts},
  year = {2024},
  eprint = {2410.22884},
  archivePrefix = {arXiv},
  primaryClass = {cs.CR},
  howpublished = {arXiv preprint arXiv:2410.22884}
}

@inproceedings{yu2022orca,
  author = {Gyeong-In Yu and Joo Seong Jeong and Geon-Woo Kim and Soojeong Kim and Byung-Gon Chun},
  title = {Orca: A Distributed Serving System for {Transformer-Based} Generative Models},
  booktitle = {Proc. OSDI},
  pages = {521--538},
  year = {2022}
}

@inproceedings{zhang2011predictive,
  author = {Danfeng Zhang and Aslan Askarov and Andrew C. Myers},
  title = {Predictive Mitigation of Timing Channels in Interactive Systems},
  booktitle = {Proc. ACM CCS},
  pages = {563--574},
  month = oct,
  year = {2011},
  doi = {10.1145/2046707.2046772}
}

@inproceedings{zhang2026hijackkv,
  author = {Yichi Zhang and Zhiqi Wang and Huan Zhang and Yuchen Yang},
  title = {{HijackKV}: New Threat in Position-Independent {KV} Cache Reuse},
  booktitle = {Proc. USENIX Security},
  pages = {4701--4720},
  month = aug,
  year = {2026},
  isbn = {978-1-939133-58-8}
}

@misc{zhang2026real,
  author = {Yage Zhang and Yukun Jiang and Zeyuan Chen and Michael Backes and Xinyue Shen and Yang Zhang},
  title = {Real Money, Fake Models: Deceptive Model Claims in Shadow {APIs}},
  year = {2026},
  eprint = {2603.01919},
  archivePrefix = {arXiv},
  note = {To appear in ACM CCS 2026},
  howpublished = {arXiv preprint arXiv:2603.01919}
}

@inproceedings{zheng2026inputsnatch,
  author = {Xinyao Zheng and Husheng Han and Shangyi Shi and Qiyan Fang and Zidong Du and Xing Hu and Qi Guo},
  title = {{InputSnatch}: Stealing Input in {LLM} Services via Cache-Sharing Timing Side-Channel Attacks},
  booktitle = {Proc. Int. Symp. Advanced Parallel Processing Technologies (APPT)},
  editor = {Lieven Eeckhout and Chao Li and Zhibin Yu and Avi Mendelson},
  series = {LNCS},
  volume = {16693},
  pages = {483--497},
  year = {2027},
  doi = {10.1007/978-981-92-4805-6_32}
}

@article{zuo2026privkvsched,
  author = {Tian-Yu Zuo and Xiuyi Zhao and Pan Li and Junxiong Lin and Mingyi Liu and Jiuchuan Jiang and Gengyu Lyu and Kai Di and Ning Luo},
  title = {{PrivKV-Sched}: Budgeted Cross-Tenant {KV}-Cache Reuse for Multi-Tenant {LLM} Serving},
  journal = {IEEE Trans. Netw. Service Manag.},
  year = {2026},
  note = {Early Access},
  doi = {10.1109/TNSM.2026.3745203}
}

\appendices
\section{Threat Models of Published Defenses}\label{app:defense-tm}

Table~\ref{tab:quotes} quotes, for each LLM-cache defense we found, what its paper lets the attacker observe; we read the published version where one exists, quotes are verbatim up to hyphenation and ligatures, and the Portuguese one is our translation.
As Section~\ref{sec:related} notes, some mention interference, ordering or all observed latency, but no threat model or evaluation asks where a hit's saving goes.

\begin{table*}[t]
\centering
\caption{What each LLM-cache defense lets the attacker observe, in its own words}
\label{tab:quotes}
\footnotesize
\begin{tabular}{@{}>{\raggedright\arraybackslash}p{0.12\textwidth}>{\raggedright\arraybackslash}p{0.66\textwidth}>{\raggedright\arraybackslash}p{0.17\textwidth}@{}}
\toprule
Defense & Quote & Where \\
\midrule
Gu et al.~\cite{gu2025auditing} & ``We assume that it is possible to send arbitrary prompts to the API \ldots\ and measure the TTFT.'' & Sec.~2.2, p.~20479 \\
PromptPeek~\cite{wu2025know} & ``This phase involves strategically sending dummy and candidate requests and observing the return order as side channel information to determine if any token matches.'' & Sec.~IV-A, p.~5 \\
Early Bird~\cite{song2025early} & ``PSA recovers the request token by token by observing per-token timing differences'' & Sec.~V-A, p.~11442 \\
InputSnatch~\cite{zheng2026inputsnatch} & ``The only required observations are request submission time and first-token or response latency.'' & Sec.~2.2, p.~486 \\
SafeKV~\cite{chu2025selective} & ``The attacker observes TTFT and aims to infer whether a victim's prefix has been cached through cross-tenant KV reuse.'' & Sec.~7, p.~8 \\
CachePrune~\cite{wu2026cacheprune} & ``We model the adversary's observation as a binary reuse oracle \ldots\ where $R(p) = 1$ indicates that a probe prompt $p$ triggers reuse of cached KV entries'' & Sec.~3.1, p.~4 \\
KVGov~\cite{addagada2026governing} & The attacker can ``Observe the TTFT of their own requests with millisecond precision'' but not ``Observe other tenants' TTFT values directly.'' & Def.~1, pp.~2--3 \\
Cache\-Solidarity~\cite{pennas2026cachesolidarity} & ``For an issued request r, the attacker atk can observe whether the prefixes associated with the request result in a sequence of hits \ldots\ and/or misses'' & Sec.~4.1, p.~7 \\
SpliceDefense~\cite{sun2026agent} & Attackers ``submit textual prompts \ldots\ and observe the end-to-end response latency'' & Sec.~IV-B, p.~6 \\
Pang et al.~\cite{pang2024cache} & ``the latency observed by the attacker should be independent of the victim's prefix cache status''; in the evaluation, ``the attacker, acting as a new user in another process, input six groups of prompts and recorded TTFT.'' & Secs.~II-C, IV-B, pp.~1239, 1241 \\
PrivKV-Sched~\cite{zuo2026privkvsched} & ``Tenants observe their own outputs, timing, and execution order and may submit candidate prompts repeatedly to test cached content.'' & Sec.~III-A, p.~3 \\
Fonseca et al.~\cite{fonseca2026mitigacao} & ``The proposal seeks to prevent an attacker from distinguishing cache hits and cache misses from the latency observed in the server's responses.'' & Sec.~1, p.~739 \\
\bottomrule
\end{tabular}
\end{table*}

\section{Proofs}\label{app:proofs}

\noindent\textbf{Lemma~\ref{lem:copy}.}
Admission is first come, first served under a per-step budget, so each step computes the next $B$ tokens of the attacker's queue that still need work.
In~$H$ the probe's $h$ cached tokens need none, so after every step the queue position reached in~$H$ is $h$ tokens ahead of~$M$ until the engine idles; since every request outputs one token, no later work absorbs this lead, which gives~(i).
If the probe runs alone in its step, every later step holds the same requests with the same shapes in both worlds, so by the shape assumption of Section~\ref{sec:system} only the probe's step differs, giving~(ii).
In~(iii), the probe and the first listener fill one step in~$H$; in~$M$ the $h$ extra tokens push the first listener's last $h$ tokens into the next step, and since every later request is $B$ long, the same $h$ tokens carry over at every step.

\noindent\textbf{Theorem~\ref{thm:ruler}.}
Besides the window conditions and in-order entry of the ruler's segments, the theorem needs a head that keeps every step after the ruler starts saturated, at least $(m+1)B/\ell + 1$ listeners before the probe and $B/\ell + 1$ after it, a tail of at least $B$ tokens, room under the sequence limit, and first-token timestamps within $\theta/2$ of their step's release, full steps being released more than $2\theta$ apart.
First, every step after the first that serves the ruler is full and gives it $B - D_k$ tokens in queue order, where $D_k$, the decodes of requests admitted before the ruler, does not increase.
Second, consecutive listeners fall in different groups exactly when a boundary lies within the $\ell$ tokens of the later one, which locates every boundary within $\ell$.
Third, at most $F^*$ of the decodes in the first step read leave before the last step read, so every step read holds between $\hat a - \tau$ and $\hat a + \tau$ of the ruler's tokens, and at most $n_{\max}$ steps lie between the two boundaries; summing gives~(i), and separation makes the two unions disjoint.
For~(iii), if release gaps of full steps lie in a known range $[d_{\mathrm{lo}}, d_{\mathrm{hi}}]$ with $(n+1)d_{\mathrm{lo}} - n d_{\mathrm{hi}} > 2\theta$, the time between the two groups reveals~$n$.
The advantage is then at least $\Pr_H[G_H] + \Pr_M[G_M] - 1$, and the two terms are equal when the conditions depend only on history before the probe is scheduled.

\noindent\textbf{Proposition~\ref{prop:membership}.}
Couple the two worlds with the same randomness: as long as every cache lookup returns the same result, the mechanism, the step shapes, every response delivered to accounts other than the victim and every account's next request coincide, so the views can first differ at a lookup that touches a block whose content differs.
Moving from one candidate to another, each differing segment is replaced first by fresh tokens and then by its target, at most $2d$ neighboring pairs that each change the view by at most~$\varepsilon$; under $(\beta,\varepsilon)$-security two steps suffice.
With $Q_i$ the view under candidate~$i$ and $i^*$ the likeliest, the best guess succeeds with probability $\sum_v \max_i \mu_i Q_i(v) \le \max\mu + (1-\max\mu)\,T$.

\noindent\textbf{Proposition~\ref{prop:burst}.}
A Poisson process has no arrival in a window of length~$w$ with probability $e^{-\lambda w}$, and solving $\lambda w \le \ln(1/p)$ gives the rate.
For a Poisson process with random intensity of mean~$\lambda$, Jensen's inequality gives at least $e^{-\lambda w}$, and for any arrival process of mean rate~$\lambda$ independent of the attacker, Markov's inequality gives at least $1 - \lambda w$, so $r_s \ge (n-1)/\bigl((1-p)/\lambda - w_0\bigr)$ suffices.

\noindent\textbf{Proposition~\ref{prop:dilution}.}
Routing that ignores content treats the honest request and the probe alike, so each reaches the instance holding~$S$ with probability~$\pi$; the honest request then saves $h$ tokens and the probe meets the single-instance conditions, and otherwise the probe's instance holds the same content in both worlds.
For uniform routing, $i$ independent requests all miss that instance with probability $(1-1/N)^i$.

\noindent\textbf{Hiding the hit request, Section~\ref{sec:hidehit}.}
If scheduling is unchanged and every other response~$Y$ is released at $\max(c_Y, t_Y)$, with readiness~$t_Y$ and $c_Y$ independent of the world, Lemma~\ref{lem:copy} gives $t_Y^H \le t_Y^M$ on every path and $<$ on an event of positive probability, so the release is no later in~$H$ and earlier with positive probability whenever $\Pr[c_Y < t_Y^M] > 0$ there.

\noindent\textbf{Theorem~\ref{thm:delay}.}
The two groups' readiness differs by~$s$ more in~$M$, engine noise is common to both worlds, and delays are independent of everything.
In~(i) the difference of group means is Gaussian with variance $\sigma_e^2 + 2\sigma^2/k$ and mean shifted by~$s$; in~(ii) the earliest of $k$ exponentials of mean~$\mu$ is exponential of mean $\mu/k$, the difference of two is Laplace with scale $\mu/k$, and such a law and its shift by~$s$ differ by $1 - e^{-ks/(2\mu)}$ in total variation.
The view contains the statistic, so the advantage is at least these distances.
For delays forming a stationary Gaussian process over the listeners' arrival order, with $k = 2^J$ listeners on each side and negligible engine noise, weighted dyadic contrasts and the Cauchy--Schwarz inequality show that an advantage below $\varepsilon_0$ requires $\sigma \ge \log_2(2k)\, s/(8\pi z)$.

\noindent\textbf{Predictive mitigation, Section~\ref{sec:inside}.}
Let every event be released on a schedule that depends only on inputs the attacker sees and on the record of mispredictions, and let $A$ be the event that none occurs in the epoch.
On~$A$ the attacker's view is the same function of its own randomness in both worlds, so its advantage is at most $\Pr_H[A^c] + \Pr_M[A^c]$, and the same sum bounds how much earlier an event can be released in~$H$ than it is ready in~$M$.

\noindent\textbf{Theorem~\ref{thm:benefit}.}
List the $V$ accounts other than~$b$ that share content with another account, and replace their shared texts one account at a time.
Each step is a neighboring pair of Definition~\ref{def:security} with that account as victim and $\{b\}$ as the observer, the other accounts sending their real traffic, so it changes $b$'s view by at most~$\varepsilon$; after the last step no content is shared across accounts, and the triangle inequality gives~(i).
For~(ii), replace~$x$ one holder at a time, while it is still shared, which is a segment-neighboring pair each time.
A utility with values in an interval of length~$R$ changes in expectation by at most $R$ times the total variation.

\noindent\textbf{Proposition~\ref{prop:lockstep}.}
The replica reads lengths, parameters, arrival times, the account partition and self-hits, never cross-account matches, so by induction over events its schedule and every release time are the same in both worlds, and honest accounts, whose responses arrive at those times with the same content, send the same requests.
Each real step computes a subset of the replica's step and, by the premise, ends within~$D$; the real cache holds every block the replica does; and with output tokens that depend only on their own request, every field of every view is the same in both worlds.

\noindent\textbf{Theorem~\ref{thm:padding}.}
A block~$c$ is a potential hit for an event~$e$ if another account that meets the victim conditions would have written it by an earlier request with the same prefix, it would still be cached, and its step's shape has not run before~$e$; mechanisms run the replica's batches or their own with the prefill budget saturated, and each shape's duration is known only between two values that the duration models take independently.
First, replace every shared text one segment at a time; by Definition~\ref{def:security} this leaves $e$'s release distribution unchanged and removes every actual cross-tenant hit.
Second, for a potential hit~$c$ append, after~$e$, a request of a third account that copies the writer's prefix, and replace that prefix by the content~$c$ would read; each move is a neighboring pair or leaves $e$'s release unchanged by causality, and at the end $c$ is a real hit whose step shape never ran before~$e$.
Two duration models that differ only on that shape are then indistinguishable to the mechanism before~$e$, so $e$'s release has the same distribution under both.
Flipping shapes one at a time from the fastest to the slowest model, the release stays the same while the time at which $e$ would finish without sharing grows by the sum of the widths, each step counted once.

\section{Engine Survey}\label{app:engines}

We read the scheduler code or documentation on the main branch of each engine between 22 and 26 October 2026.
Our experiments use vLLM~0.26, SGLang~0.5.16 and, for CacheSolidarity, the authors' vLLM~0.8.5.
Engines release outputs per step, but clients may see coarser groups because of stream intervals, stop-string buffers, merging when a client lags, or speculative decoding; this coarsens what the attacker sees but not the step-level argument.
A gateway that buffers whole responses holds even their headers until they end, which for listeners that return one token barely moves their first token.

\section{Additional Experiments}\label{app:experiments}

\noindent\textbf{Setup.}
GPUs are L40S with 48\,GB, RTX PRO 6000 with 96\,GB and H100 NVL with 94\,GB; engines run with their default schedulers, a context limit of 16,384 tokens, or 32,768 on RTX PRO 6000 and H100 GPUs, 85\% of GPU memory and seed~0.
Models are Llama-3.1-8B-Instruct, Qwen3-8B, Llama-3.1-70B in FP8 over two GPUs, DeepSeek-V2-Lite, a mixture of experts with 2.4B of its 15.7B parameters active per token, on one GPU without expert parallelism, and Qwen3.5-9B for the hybrid case below.
The router is the vLLM production stack at commit \texttt{7b1164db}.
Each trial is one cache epoch with fresh random tokens, its world drawn with probability one half and its cell, the construction and secret length, drawn at random; honest load replays the conversation part of the Azure trace of 2023 at fractions of the rate that saturates the engine.
BurstGPT's six busiest consecutive hours in its first file hold 99,345 requests with timestamps in whole seconds, each offset uniformly within its second; its calibration stopped at half its original rate, where bursts lengthened latency, so we also replayed it at its original rate and at twice that.

\noindent\textbf{Sample sizes.}
Table~\ref{tab:samples} covers Sections~\ref{sec:attack} and~\ref{sec:defenses}.

\begin{table}[t]
\centering
\caption{Trials per world and intervals}
\label{tab:samples}
\footnotesize
\setlength{\tabcolsep}{4pt}
\begin{tabular}{@{}>{\raggedright\arraybackslash}p{0.36\columnwidth}rr>{\raggedright\arraybackslash}p{0.21\columnwidth}@{}}
\toprule
Result & Trials & Advantage & Interval \\
\midrule
Copied saving, vLLM & 132--172 & 0.985--0.987 & L $\ge$ 0.977 \\
Same, SGLang & 81--125 & 0.957--0.983 & L $\ge$ 0.938 \\
Same, Qwen3-8B & 90--107 & 0.980--0.983 & L $\ge$ 0.964 \\
Four requests & 133--166 & 0.985--0.989 & groups disjoint \\
Ruler, vLLM, to half load & 37--66 & 0.980--1.000 & B90 $\ge$ 0.941 \\
Ruler, SGLang, to a quarter & 32--48 & 0.963--1.000 & B90 $\ge$ 0.926 \\
Two instances, affinity & 47--55 & 0.867--0.980 & B90 $\ge$ 0.791 \\
Four instances, affinity & 42--55 & 0.599--0.980 & B90 $\ge$ 0.471 \\
Round robin, a probe each & 30--50 & 0.344--0.844 & B90 $\ge$ 0.189 \\
RTX PRO 6000, affinity & 43--60 & 0.754--0.933 & B90 $\ge$ 0.629 \\
70B, to a quarter & 31--55 & 0.982--1.000 & B90 $\ge$ 0.945 \\
H100, to a quarter & 55--77 & 0.982--1.000 & B90 $\ge$ 0.964 \\
BurstGPT trace & 34--61 & 0.980--1.000 & B90 $\ge$ 0.951 \\
Mixture of experts, to half load & 38--68 & 0.948--1.000 & B90 $\ge$ 0.916 \\
TLS proxy, other node & 46--70 & 0.882--0.975 & B90 $\ge$ 0.827 \\
End-to-end, accuracy & 120 & 0.851--0.971 & B90 $\ge$ 0.804 \\
KVGov noise, to half load & 40--66 & 0.987--1.000 & B90 $\ge$ 0.965 \\
CacheSolidarity, fixed & 83--115 & 0.973--0.981 & groups disjoint \\
\bottomrule
\end{tabular}
\tablenote{For the results of Sections~\ref{sec:attack} and~\ref{sec:defenses}.
L: one-sided 95\% bootstrap lower bound of the cross-fitted advantage; B90: 90\% percentile bootstrap interval of the decision rule's advantage; groups disjoint: the two worlds' readings do not overlap.}
\end{table}

\noindent\textbf{Burst rates.}
Table~\ref{tab:burst} gives the burst rate that Proposition~\ref{prop:burst} requires for $w_0 = 22$\,ms with the rulers of Section~\ref{sec:ruler}, which send 19, 51 and 175 requests in the window for secrets of 4,096, 1,024 and 256 tokens.
For any arrival process of the same mean rate, the sufficient rate replaces $\ln(1/p)$ by $1-p$; for $p = 0.8$ it is 48, 132 and 460 per second at 0.5 honest requests per second, and 272, 756 and 2,632 at 2.27.
No rate suffices once $\lambda w_0 \ge \ln(1/p)$: for $\lambda$ above 31.5, 10.1 and 4.8 per second at $p$ of 0.5, 0.8 and 0.9.

\begin{table}[t]
\centering
\caption{Required burst rate, in requests per second}
\label{tab:burst}
\footnotesize
\begin{tabular}{@{}rrrrr@{}}
\toprule
$\lambda$ & $p$ & 4,096 & 1,024 & 256 \\
\midrule
0.5 & 0.5 & 13 & 37 & 128 \\
0.5 & 0.8 & 42 & 118 & 410 \\
0.5 & 0.9 & 95 & 265 & 922 \\
2.27 & 0.5 & 64 & 176 & 614 \\
2.27 & 0.8 & 236 & 655 & 2,280 \\
2.27 & 0.9 & 737 & 2,048 & 7,127 \\
\bottomrule
\end{tabular}
\tablenote{The burst rate for which no honest Poisson request enters the window with probability at least~$p$, by honest rate~$\lambda$ per second and secret length in tokens; rounded.}
\end{table}

\noindent\textbf{Four requests.}
In an analysis made after seeing the results, the probe's own first token arrived 1.4 to 4.4\,ms earlier in~$M$ for hits up to 1,024 tokens: its step holds $B$ tokens in both worlds and was shorter in~$M$, for reasons not isolated, so delaying such a hit cannot release it at its miss-world time.

\noindent\textbf{Several API servers.}
With two API servers in front of one engine at half of capacity, the attacker's own requests entered out of order and the 1,024-token ruler fell to advantage 0.38; spacing its groups by 30\,ms restored 0.93.

\noindent\textbf{Gateway.}
Caddy~2.8.4 terminated TLS on the GPU's node with a self-signed certificate and forwarded to vLLM; the attacker, the honest traffic and the metrics poll ran on a CPU node of the same cluster.

\noindent\textbf{Measuring the budget.}
On an idle engine, the attacker sends a burst of equal requests of $\ell$ fresh tokens and groups their first tokens at gaps above 5\,ms.
The estimator fixed before the first run, from mean group sizes, fell short: vLLM processes about 128 outputs at a time, so a step with more requests is released in several groups, and asynchronous scheduling splits the sequence limit between two batches in flight.
A second estimator, fixed before a second run, takes $\ell$ times the median group size, with the smallest $\ell$ among 64, 100 and 160 whose median group is at most 100.
It gave the budget within $\ell$ in all 8 bursts for budgets of 2,048, 4,096 and 8,192 and with the sequence limit set to 64; the sum of two consecutive groups gave the sequence limit in 6 to 8 of 8.

\noindent\textbf{Defenses.}
Predictive mitigation doubled its penalty after each misprediction of any of the attacker's requests, with a predictor fitted per cell and listener on a third of the miss-world trials, which were not scored, and one-sided 95\% Clopper--Pearson bounds.
Margins of timed steps are nominal and not comparable across runs: the calibrated steps of the first run and of the repetition were 2.6\% and 1.2\% faster than back-to-back miss-world steps.
The first run's 5\% deadline, 1.023 back-to-back steps, therefore lay between the repetition's 3\% and 4\%, at 1.017 and 1.027; the runs had 100 and 80 trials per world and margin.

\noindent\textbf{CacheSolidarity.}
We ran the authors' code at its public commit on its own vLLM~0.8.5 with the kernels from their published image, with its block log on a local memory disk.
A reused cache block keeps its flag though its hash is cleared, so stale flags cut unrelated hits; as runs proceed, most cross-tenant hits are cut in both worlds and multi-block probes fall to 0.22 to 0.46, while one-block hits still read 0.977.
This defect drops sharing at random and is no protection; our one-line fix clears the flag with the hash, and Section~\ref{sec:narrow} evaluates the fixed code.
With the defense off, flags and decoys read at most 0.1 and the ruler 0.972.

\noindent\textbf{Hybrid models.}
On Qwen3.5-9B, which mixes full and linear attention, the full-attention cache can hold a longer prefix of a request than the linear-state cache, as after another account cached a shared prefix.
vLLM then ends the prefill chunk there to store the linear state, so a request that hits nothing takes 11 to 25\,ms longer, and the worlds do not overlap.

\end{document}